\documentclass[aps,prd,twocolumn,superscriptaddress]{revtex4-1}

\usepackage{times}
\usepackage{graphicx,psfrag,units}
\usepackage{amsmath,amsfonts,amssymb}
\usepackage{microtype}
\usepackage{color}
\usepackage{multirow}

\usepackage[bookmarksnumbered]{hyperref}
\usepackage[all]{hypcap}

\def\thercsid{\relax}
\def\rcsid#1{\def\next##1#1{\def\thercsid{##1}}\next}
\rcsid$Rev: 10206 $

\begin{document}

\title{Searching for stochastic gravitational waves using data from \\
the two co-located LIGO Hanford detectors} ~\\


\author{%
J.~Aasi,$^{1}$
J.~Abadie,$^{1}$
B.~P.~Abbott,$^{1}$
R.~Abbott,$^{1}$
T.~Abbott,$^{2}$
M.~R.~Abernathy,$^{1}$
T.~Accadia,$^{3}$
F.~Acernese,$^{4,5}$
C.~Adams,$^{6}$
T.~Adams,$^{7}$
% ...
}\noaffiliation

\affiliation {LIGO - California Institute of Technology, Pasadena, CA 91125, USA }
\affiliation {Louisiana State University, Baton Rouge, LA 70803, USA }
\affiliation {Laboratoire d'Annecy-le-Vieux de Physique des Particules (LAPP), Universit\'e de Savoie, CNRS/IN2P3, F-74941 Annecy-le-Vieux, France }
\affiliation {Universit\`a di Salerno, Fisciano, I-84084 Salerno, Italy }
\affiliation {INFN, Sezione di Napoli, Complesso Universitario di Monte S.Angelo, I-80126 Napoli, Italy }
\affiliation {LIGO - Livingston Observatory, Livingston, LA 70754, USA }
\affiliation {Cardiff University, Cardiff, CF24 3AA, United Kingdom }
\affiliation {University of Salerno, I-84084 Fisciano (Salerno), Italy and INFN, Italy }
\affiliation {Albert-Einstein-Institut, Max-Planck-Institut f\"ur Gravitationsphysik, D-30167 Hannover, Germany }
\affiliation {Nikhef, Science Park, 1098 XG Amsterdam, The Netherlands }
\affiliation {LIGO - Massachusetts Institute of Technology, Cambridge, MA 02139, USA }
\affiliation {Instituto Nacional de Pesquisas Espaciais, 12227-010 - S\~{a}o Jos\'{e} dos Campos, SP, Brazil }
\affiliation {University of Wisconsin--Milwaukee, Milwaukee, WI 53201, USA }
\affiliation {Leibniz Universit\"at Hannover, D-30167 Hannover, Germany }
\affiliation {Universit\`a di Siena, I-53100 Siena, Italy }
\affiliation {INFN, Sezione di Pisa, I-56127 Pisa, Italy }
\affiliation {University of Florida, Gainesville, FL 32611, USA }
\affiliation {The University of Mississippi, University, MS 38677, USA }
\affiliation {California State University Fullerton, Fullerton, CA 92831, USA }
\affiliation {INFN, Sezione di Roma, I-00185 Roma, Italy }
\affiliation {University of Birmingham, Birmingham, B15 2TT, United Kingdom }
\affiliation {Albert-Einstein-Institut, Max-Planck-Institut f\"ur Gravitationsphysik, D-14476 Golm, Germany }
\affiliation {Montana State University, Bozeman, MT 59717, USA }
\affiliation {European Gravitational Observatory (EGO), I-56021 Cascina, Pisa, Italy }
\affiliation {Syracuse University, Syracuse, NY 13244, USA }
\affiliation {LIGO - Hanford Observatory, Richland, WA 99352, USA }
\affiliation {SUPA, University of Glasgow, Glasgow, G12 8QQ, United Kingdom }
\affiliation {APC, AstroParticule et Cosmologie, Universit\'e Paris Diderot, CNRS/IN2P3, CEA/Irfu, Observatoire de Paris, Sorbonne Paris Cit\'e, 10, rue Alice Domon et L\'eonie Duquet, F-75205 Paris Cedex 13, France }
\affiliation {Columbia University, New York, NY 10027, USA }
\affiliation {Stanford University, Stanford, CA 94305, USA }
\affiliation {Universit\`a di Pisa, I-56127 Pisa, Italy }
\affiliation {CAMK-PAN, 00-716 Warsaw, Poland }
\affiliation {Universit\`a degli Studi di Genova, I-16146 Genova, Italy }
\affiliation {INFN, Sezione di Genova, I-16146 Genova, Italy }
\affiliation {The University of Texas at Brownsville, Brownsville, TX 78520, USA }
\affiliation {San Jose State University, San Jose, CA 95192, USA }
\affiliation {Moscow State University, Moscow, 119992, Russia }
\affiliation {LAL, Universit\'e Paris-Sud, IN2P3/CNRS, F-91898 Orsay, France }
\affiliation {NASA/Goddard Space Flight Center, Greenbelt, MD 20771, USA }
\affiliation {University of Western Australia, Crawley, WA 6009, Australia }
\affiliation {ARTEMIS, Universit\'e Nice-Sophia-Antipolis, CNRS and Observatoire de la C\^ote d'Azur, F-06304 Nice, France }
\affiliation {Institut de Physique de Rennes, CNRS, Universit\'e de Rennes 1, F-35042 Rennes, France }
\affiliation {Laboratoire des Mat\'eriaux Avanc\'es (LMA), IN2P3/CNRS, Universit\'e de Lyon, F-69622 Villeurbanne, Lyon, France }
\affiliation {Washington State University, Pullman, WA 99164, USA }
\affiliation {INFN, Sezione di Perugia, I-06123 Perugia, Italy }
\affiliation {Universit\`a degli Studi di Urbino 'Carlo Bo', I-61029 Urbino, Italy }
\affiliation {INFN, Sezione di Firenze, I-50019 Sesto Fiorentino, Firenze, Italy }
\affiliation {University of Oregon, Eugene, OR 97403, USA }
\affiliation {Laboratoire Kastler Brossel, ENS, CNRS, UPMC, Universit\'e Pierre et Marie Curie, F-75005 Paris, France }
\affiliation {Astronomical Observatory Warsaw University, 00-478 Warsaw, Poland }
\affiliation {VU University Amsterdam, 1081 HV Amsterdam, The Netherlands }
\affiliation {University of Maryland, College Park, MD 20742, USA }
\affiliation {University of Massachusetts - Amherst, Amherst, MA 01003, USA }
\affiliation {Universitat de les Illes Balears, E-07122 Palma de Mallorca, Spain }
\affiliation {Universit\`a di Napoli 'Federico II', Complesso Universitario di Monte S.Angelo, I-80126 Napoli, Italy }
\affiliation {Canadian Institute for Theoretical Astrophysics, University of Toronto, Toronto, Ontario, M5S 3H8, Canada }
\affiliation {Tsinghua University, Beijing 100084, China }
\affiliation {University of Michigan, Ann Arbor, MI 48109, USA }
\affiliation {Rochester Institute of Technology, Rochester, NY 14623, USA }
\affiliation {INFN, Sezione di Roma Tor Vergata, I-00133 Roma, Italy }
\affiliation {National Tsing Hua University, Hsinchu Taiwan 300 }
\affiliation {Charles Sturt University, Wagga Wagga, NSW 2678, Australia }
\affiliation {Caltech-CaRT, Pasadena, CA 91125, USA }
\affiliation {Pusan National University, Busan 609-735, Korea }
\affiliation {Australian National University, Canberra, ACT 0200, Australia }
\affiliation {Carleton College, Northfield, MN 55057, USA }
\affiliation {Universit\`a di Roma Tor Vergata, I-00133 Roma, Italy }
\affiliation {Universit\`a di Roma 'La Sapienza', I-00185 Roma, Italy }
\affiliation {The George Washington University, Washington, DC 20052, USA }
\affiliation {University of Cambridge, Cambridge, CB2 1TN, United Kingdom }
\affiliation {University of Minnesota, Minneapolis, MN 55455, USA }
\affiliation {The University of Sheffield, Sheffield S10 2TN, United Kingdom }
\affiliation {Wigner RCP, RMKI, H-1121 Budapest, Konkoly Thege Mikl\'os \'ut 29-33, Hungary }
\affiliation {University of Sannio at Benevento, I-82100 Benevento, Italy and INFN (Sezione di Napoli), Italy }
\affiliation {Inter-University Centre for Astronomy and Astrophysics, Pune - 411007, India }
\affiliation {Universit\`a di Trento, I-38123 Povo, Trento, Italy }
\affiliation {INFN, Gruppo Collegato di Trento, I-38050 Povo, Trento, Italy }
\affiliation {California Institute of Technology, Pasadena, CA 91125, USA }
\affiliation {Northwestern University, Evanston, IL 60208, USA }
\affiliation {Montclair State University, Montclair, NJ 07043, USA }
\affiliation {The Pennsylvania State University, University Park, PA 16802, USA }
\affiliation {MTA E\"otv\"os University, `Lendulet' Astrophysics Research Group, Budapest 1117, Hungary }
\affiliation {National Astronomical Observatory of Japan, Tokyo 181-8588, Japan }
\affiliation {Universit\`a di Perugia, I-06123 Perugia, Italy }
\affiliation {University of Szeged, D\'om t\'er 9, Szeged 6720, Hungary }
\affiliation {Rutherford Appleton Laboratory, HSIC, Chilton, Didcot, Oxon, OX11 0QX, United Kingdom }
\affiliation {Embry-Riddle Aeronautical University, Prescott, AZ 86301, USA }
\affiliation {Perimeter Institute for Theoretical Physics, Ontario, N2L 2Y5, Canada }
\affiliation {American University, Washington, DC 20016, USA }
\affiliation {College of William and Mary, Williamsburg, VA 23187, USA }
\affiliation {University of Adelaide, Adelaide, SA 5005, Australia }
\affiliation {Raman Research Institute, Bangalore, Karnataka 560080, India }
\affiliation {Korea Institute of Science and Technology Information, Daejeon 305-806, Korea }
\affiliation {University of Bia{\l }ystok, 15-424 Bia{\l }ystok, Poland }
\affiliation {University of Southampton, Southampton, SO17 1BJ, United Kingdom }
\affiliation {IISER-TVM, CET Campus, Trivandrum Kerala 695016, India }
\affiliation {Hobart and William Smith Colleges, Geneva, NY 14456, USA }
\affiliation {Institute of Applied Physics, Nizhny Novgorod, 603950, Russia }
\affiliation {Seoul National University, Seoul 151-742, Korea }
\affiliation {Hanyang University, Seoul 133-791, Korea }
\affiliation {NCBJ, 05-400 \'Swierk-Otwock, Poland }
\affiliation {IM-PAN, 00-956 Warsaw, Poland }
\affiliation {Institute for Plasma Research, Bhat, Gandhinagar 382428, India }
\affiliation {Utah State University, Logan, UT 84322, USA }
\affiliation {The University of Melbourne, Parkville, VIC 3010, Australia }
\affiliation {University of Brussels, Brussels 1050 Belgium }
\affiliation {SUPA, University of Strathclyde, Glasgow, G1 1XQ, United Kingdom }
\affiliation {ESPCI, CNRS, F-75005 Paris, France }
\affiliation {Universit\`a di Camerino, Dipartimento di Fisica, I-62032 Camerino, Italy }
\affiliation {The University of Texas at Austin, Austin, TX 78712, USA }
\affiliation {Southern University and A\&M College, Baton Rouge, LA 70813, USA }
\affiliation {IISER-Kolkata, Mohanpur, West Bengal 741252, India }
\affiliation {National Institute for Mathematical Sciences, Daejeon 305-390, Korea }
\affiliation {RRCAT, Indore MP 452013, India }
\affiliation {Tata Institute for Fundamental Research, Mumbai 400005, India }
\affiliation {Louisiana Tech University, Ruston, LA 71272, USA }
\affiliation {SUPA, University of the West of Scotland, Paisley, PA1 2BE, United Kingdom }
\affiliation {Institute of Astronomy, 65-265 Zielona G\'ora, Poland }
\affiliation {Indian Institute of Technology, Gandhinagar Ahmedabad Gujarat 382424, India }
\affiliation {Department of Astrophysics/IMAPP, Radboud University Nijmegen, P.O. Box 9010, 6500 GL Nijmegen, The Netherlands }
\affiliation {Andrews University, Berrien Springs, MI 49104, USA }
\affiliation {Trinity University, San Antonio, TX 78212, USA }
\affiliation {INFN, Sezione di Padova, I-35131 Padova, Italy }
\affiliation {University of Washington, Seattle, WA 98195, USA }
\affiliation {Southeastern Louisiana University, Hammond, LA 70402, USA }
\affiliation {Abilene Christian University, Abilene, TX 79699, USA }

\date[\relax]{ RCS \thercsid; compiled \today }

\begin{abstract}\quad
Searches for a stochastic gravitational-wave background (SGWB) using
terrestrial detectors typically involve cross-correlating data from pairs of
detectors. The sensitivity of such cross-correlation analyses depends,
among other things, on the separation between the two detectors: the smaller 
the separation, the better the sensitivity. Hence, a co-located detector 
pair is more sensitive to a gravitational-wave background than a 
non-co-located detector pair. However, co-located detectors are also 
expected to suffer from correlated noise from instrumental and environmental 
effects that could contaminate the measurement of the background.
Hence, methods to identify and mitigate the effects of correlated noise are
necessary to achieve the potential increase in sensitivity of co-located 
detectors. Here we report on the first SGWB analysis using the two LIGO 
Hanford detectors and address the complications arising from correlated 
environmental noise. We apply correlated noise identification and mitigation 
techniques to data taken by the two LIGO Hanford detectors, H1 and H2, 
during LIGO's fifth science run. At low frequencies, $\unit[40 - 460] {Hz}$, 
we are unable to sufficiently mitigate the correlated noise to a level where 
we may confidently measure or bound the stochastic gravitational-wave signal.
However, at high frequencies, $\unit[460-1000]{Hz}$, these techniques are
sufficient to set a $95\%$ confidence level (C.L.) upper limit on the
gravitational-wave energy density of $\Omega(f)<7.7\times 10^{-4}
(f/\unit[900]{Hz})^3$, which improves on the previous upper limit by a
factor of $\sim 180$. In doing so, we demonstrate techniques that will be 
useful for future searches using advanced detectors, where correlated noise 
(e.g., from global magnetic fields) may affect even widely separated detectors.
\end{abstract}

\maketitle

\section{Introduction}
\label{sec:intro}

The detection of a stochastic gravitational-wave background (SGWB), of either
cosmological or astrophysical origin, is a major science goal for both current
and planned searches for gravitational waves (GWs)
\cite{Grishchuk:1976,starobinskii,barkana,maggiore}. Given the weakness of
the gravitational interaction, cosmological GWs are expected to decouple
from matter in the early universe much earlier than any other form of radiation
(e.g., photons, neutrinos, etc.). The detection of such a primordial GW
background by the current ground-based detectors \cite{LIGOdescription, Virgo,
GEO}, proposed space-based detectors \cite{LISA, BBO}, or a pulsar timing
array \cite{Jenet:2006, vanHaasteren-et-al:2011} would give us a picture of
the universe mere fractions of a second after the Big-Bang
\cite{Grishchuk:1976,starobinskii,barkana,allenleshouches},
allowing us to study the physics of the highest energy scales, unachievable in
standard laboratory experiments \cite{maggiore}. The recent results from the
BICEP2 experiment indicate the existence of cosmic microwave background
B-mode polarization at degree angular scales \cite{BICEP2}, which may be due to
an ultra-low frequency primordial GW background, such as would be generated by
amplification of vacuum fluctuations during cosmological inflation; however, it
cannot currently be ruled out that the observed B-mode polarization is due to
a Galactic dust foreground \cite{BICEP2_dust1,BICEP2_dust2}). These GWs and
their high frequency counterparts in standard slow-roll inflationary model are
several orders of magnitude below the sensitivity levels of current and
advanced LIGO detectors. Hence they are not the target of our current analysis.
However, many non-standard inflationary models predict GWs that could be
detected by advanced LIGO detectors.

On the other hand, the detection of a SGWB due to spatially and temporally 
unresolved foreground astrophysical sources such as magnetars 
\cite{magnetars}, rotating
neutron stars \cite{rotneutrstar}, galactic and extragalactic compact binaries
\cite{SGWBBNS, SGWBBNSextra, SGWBBBH},
or the inspiral and collisions of supermassive black holes associated with
distant galaxy mergers \cite{Jaffe:2003}, would provide information about the
spatial distribution and formation rate of these various source populations.

Given the random nature of a SGWB, searches require cross-correlating data 
from two or more detectors
\cite{Grishchuk:1976, Michelson:1987, ChristensenI, Flanagan, allenromano},
under the assumption that correlated noise between any two detectors is
negligible. For such a case, the contribution to the cross-correlation
from the (common) GW signal grows linearly with the observation time $T$,
while that from the noise grows like $\sqrt{T}$. Thus, the signal-to-noise 
ratio (SNR) also grows like $\sqrt{T}$. This allows one to search for 
stochastic signals buried within the detector noise by integrating for 
a sufficiently long interval of time.

For the widely-separated detectors in Livingston, LA and Hanford, WA, the
physical separation
($\sim\!3000~{\rm km}$) eliminates the coupling of local instrumental and
environmental noise between the two detectors, while global disturbances
such as electromagnetic resonances are at a sufficiently low
level that they are not observable in coherence measurements between
the (first-generation) detectors at their design
sensitivity~\cite{LIGOdescription, S1HLiso, S4HLiso, S5HLiso,schumann, wsubtract}.

While physically-separated detectors have the advantage of reduced correlated
noise, they have the disadvantage of reduced sensitivity to a SGWB;
physically-separated detectors respond at different times to GWs from 
different directions and with differing response amplitudes
depending on the relative orientation and (mis)alignment of the detectors
\cite{ChristensenI, Flanagan, allenromano}. Co-located and co-aligned
detectors, on the other hand, such as the 4~km and 2~km interferometers in
Hanford, WA (denoted H1 and H2), respond identically to GWs from all
directions and for all frequencies below a few kHz. They are thus, potentially,
an order-of-magnitude more sensitive to a SGWB than e.g., the 
Hanford-Livingston LIGO pair. But this potential gain in
sensitivity can be offset by the presence of correlated instrumental and
environmental noise, given that the two detectors share the same local
environment. Methods to identify and mitigate the effects of correlated
noise are thus needed to realize the potential increase in sensitivity
of co-located detectors.

In this paper, we apply several noise identification and mitigation techniques
to data taken by the two LIGO Hanford detectors, H1 and H2, during LIGO's 
fifth science run (S5, November 4, 2005, to September 30, 2007) in the 
context of a search for a SGWB. This is the first stochastic analysis 
using LIGO science data that addresses the complications introduced by
correlated environmental noise. As discussed in the references
\cite{schumann, wsubtract}, the coupling of global magnetic fields to
non-colocated advanced LIGO detectors could produce significant correlations
between them thereby reducing their sensitivity
to SGWB by an order of magnitude. We expect the current H1-H2 analysis to
provide a useful precedent for SGWB searches with advanced detectors in such
(expected) correlated noise environment.

Results are presented at different stages of cleaning applied to the data.
We split the analysis into two parts---one for the frequency band 
460--1000~Hz, where we are able to successfully identify and exclude 
significant narrow-band correlations; and the other for the band 
80--160~Hz, where even after applying the noise reduction methods there 
is still evidence of residual contamination, resulting
in a large systematic uncertainty for this band. The frequencies below 80~Hz
and between 160--460~Hz are not included in the analysis because of poor
detector sensitivity and contamination by known noise artifacts.
We observe no evidence of a SGWB and so our final results are given in the 
form of upper-limits. Due to the presence of residual correlated noise between
80--160~Hz, we do not set any upper-limit for this frequency band. Since we 
do not observe any such residual noise between 460--1000~Hz, in that 
frequency band and the 5 sub-bands assigned to it, we set astrophysical 
upper-limits on the energy density of stochastic GWs.

The rest of the paper is organized as follows. In Sec.~\ref{sec:noise}, we
describe sources of correlated noise in H1 and H2, and the environmental and
instrumental monitoring system. In Sec.~\ref{sec:cross-correlation} we
describe the cross-correlation procedure used to search for a SGWB.
In Secs.~\ref{sec:methods} and \ref{sec:steps} we describe the
methods that we used to identify correlated noise, and the steps that we
took to mitigate it. In Secs.~\ref{sec:results} and \ref{sec:upperlimits}
we give the results of our analysis applied to the S5 H1-H2 data. Finally, in
Sec.~\ref{sec:summary} we summarize our results and discuss potential
improvements to the methods discussed in this paper.

\section{Common noise in the two LIGO Hanford detectors}
\label{sec:noise}

At each of the LIGO observatory sites the detectors are supplemented
with a set of sensors to monitor the local environment~\cite{S1,
LIGOdescription}. Seismometers and accelerometers measure vibrations of
the ground and various detector components; microphones monitor
acoustic noise; magnetometers monitor magnetic fields that could couple to 
the test masses (end mirrors of the interferometers) via the magnets 
attached to the test masses to control their positions; radio receivers 
monitor radio frequency (RF) power around the laser modulation frequencies, 
and voltage line monitors record fluctuations in the AC power. These physical 
environment monitoring (PEM) channels are used to detect instrumental and 
environmental disturbances that can couple to the GW strain channel. We 
assume that these channels are completely insensitive to GW strain. The PEM 
channels are placed at strategic locations around the observatory, especially 
near the corner and ends of the L-shaped interferometer where important 
laser, optical, and suspension systems reside in addition to the test 
masses themselves.

Information provided by the PEM channels is used in many different ways. The
most basic application is the creation of numerous {\em data quality flags}
identifying stretches of data that are corrupted by instrumental or
environmental noise~\cite{S5glitch}. The signals from PEM channels are
critical in defining these flags; microphones register airplanes flying 
overhead, seismometers and accelerometers detect elevated seismic activity or
anthropogenic events (trucks, trains, logging), and magnetometers detect 
fluctuations in the mains power supply and the Earth's magnetic field.

In searches for transient GW signals, such as burst or coalescing binary
events, information from the PEM channels has been used to construct
\textit{vetoes}~\cite{ChristensenVeto1, BurstVeto1, ChristensenVeto2, S4burst}.
When a clear association can be made between a measured environmental
event and a coincident \textit{glitch} in the output channel of the
detector, then these times are excluded from the transient GW searches.
These event-by-event vetoes exclude times of order hundreds of milliseconds
to a few seconds.

Similarly, noise at specific frequencies, called \textit{noise lines}, can
affect searches for GWs from rotating neutron stars or even for a SGWB.
In S5, data from PEM channels were used to verify that some of the
apparent periodic signals were in fact due to noise sources at the
observatories~\cite{S5pulsarPF, S5pulsarEinstein}. Typically the
neutron-star search algorithms can also be applied to the PEM data to
find channels that have noise lines at the same frequencies as those in the
detector output channel. The coherence is also calculated between the
detector output and the PEM channels, and these results provide additional
information for determining the source of noise lines.

The study of noise lines has also benefited past LIGO searches for stochastic
GWs. For example, in LIGO's search for a SGWB using the data from the 
S4 run~\cite{S4HLiso}, correlated noise between the Hanford and Livingston 
detectors was observed in the form of a forest of sharp 1~Hz harmonic lines.
It was subsequently determined that these lines were caused by the sharp 
ramp of a one-pulse-per-second signal, injected into the data acquisition 
system to synchronize it with the Global Positioning System (GPS) time 
reference. In the S5 stochastic search~\cite{S5HLiso}, there were other 
prominent noise lines that were subsequently identified through the use of 
the PEM signals.

In addition to \textit{passive} studies, where the PEM signals are
observed and associations are made to detector noise, there have also
been a series of \textit{active} investigations where noise was injected
into the detector environment in order to measure its coupling
to the GW channel. Acoustic, seismic, magnetic, and RF electromagnetic noise 
were injected into the observatory environment at various locations and 
responses of the detectors were studied. These tests provided clues and ways 
to better isolate the detectors from the environment.

All the previous LIGO searches for a SGWB have used the
physically-separated Hanford and Livingston detectors and assumed that
common noise between these non-colocated detectors was inconsequential.
This assumption was strongly supported by observations---i.e., none of the
coherence measurements performed to date between these detectors
revealed the presence of correlations other than those known to be introduced
by the instrument itself (for example, harmonics of the 60~Hz power line).
Since the analysis presented here uses the two co-located Hanford detectors,
which are susceptible to correlated noise due to the local environment, new
methods were required to identify and mitigate the correlated noise.

\section{cross-correlation procedure}
\label{sec:cross-correlation}
The energy density spectrum of SGWB is defined as
\begin{equation}
  \Omega_{\rm gw}(f) \equiv \frac{f}{\rho_c}\frac{d\rho_{\rm gw}}{df}
\end{equation}
where $\rho_c$ ($=\frac{3c^2H_0^2}{8 \pi G}$) is the critical energy density
and $\rho_{\rm gw}$ is the GW energy density contained in the frequency range
$f$ and $f+df$.
Since most theoretical models of stochastic backgrounds in the LIGO band are
characterized by a power-law spectrum, we will assume that the fractional
energy density in GWs \cite{Grishchuk:1988} has the form
\begin{equation}
\Omega_{\rm gw}(f) = \Omega_\alpha
\left(\frac{f}{f_{\rm ref}}\right)^\alpha\,,
\label{e:Omega_gw_alpha}
\end{equation}
where $\alpha$ is the spectral index and $f_{\rm ref}$ is some reference
frequency.
We will consider two values for the spectral index: $\alpha=0$ which is
representative of many cosmological models, and $\alpha=3$ which is
characteristic of many astrophysical models. This latter case corresponds to
a flat (i.e., constant) one-sided power spectral density (PSD) in the strain
output of a detector $S_{\rm gw}(f)$, since
\begin{equation}
S_{\rm gw}(f) =
\frac{3 H_0^2}{10\pi^2}
\frac{\Omega_{\rm gw}(f)}{f^3}
\propto f^{\alpha-3}\,.
\label{e:Sgw}
\end{equation}
Here $H_0$ is the present value of the Hubble parameter, assumed to
be $H_0 = 68$ km/s/Mpc \cite{hubble}.

Following the procedures described in \cite{allenromano}, we construct
our cross-correlation statistic as estimators of
$\Omega_\alpha$ for {\em individual} frequency bins, of width $\Delta f$,
centered at each (positive) frequency $f$. These estimators are simply the
measured values of the cross-spectrum of the strain output of two detectors
divided by the expected shape of the cross-correlation due to a GW
background with spectral index $\alpha$:
\begin{align}
\hat\Omega_\alpha(f)
\equiv
\frac{2}{T}
\frac{\Re\left[\tilde s_1^*(f)\tilde s_2(f)\right]}
{\gamma(f) S_{\alpha}(f)}\,.
\label{e:Omega_est}
\end{align}
Here $T$ is the duration of the data segments used for Fourier transforms;
$\tilde s_1(f)$, $\tilde s_2(f)$ are the Fourier transforms of the strain
time-series in the two detectors; $S_\alpha(f)$ is proportional to the assumed
spectral shape,
\begin{equation}
S_\alpha(f)
\equiv \frac{3 H_0^2}{10\pi^2}
\frac{1}{f^3}
\left(\frac{f}{f_{\rm ref}}\right)^\alpha\,;
\label{e:Salpha}
\end{equation}
and $\gamma(f)$ is the overlap reduction function
\cite{ChristensenI, Flanagan, allenromano}, which encodes the reduction in
sensitivity due to the separation and relative alignment of the two detectors.
For the H1-H2 detector pair, $\gamma(f) \approx 1$ for all frequencies below a
few kHz\footnote{For other pairs such as H1-L1, $\gamma(f)$ could be zero
at some frequencies. Since
both $\hat\Omega_\alpha(f)$ and $\sigma_{\hat\Omega_\alpha}(f)$ have
$\gamma(f)$ in their denominator, the final result obtained by the
weighted average (ref Eq.\ref{e:summed_estimator}) is always finite.}.

In the absence of correlated noise, one can show that the above estimators
are {\em optimal}---i.e., they are unbiased, minimal-variance estimators of
$\Omega_{\alpha}$ for stochastic background signals with spectral index
$\alpha$. Assuming that the detector noise is Gaussian, stationary, and
much larger in magnitude than the GW signal, the expectation value of the
variance of the estimators is given by
\begin{align}
\sigma^2_{\hat\Omega_\alpha}(f)
\approx \frac{1}{2T\Delta f}
\frac{P_1(f)P_2(f)}{\gamma^2(f)S_\alpha^2(f)}\,,
\label{e:var_Omega_est}
\end{align}
where $P_1(f)$, $P_2(f)$ are the one-sided PSDs of the detector output
$\tilde s_1(f)$, $\tilde s_2(f)$ respectively. For a frequency band consisting of several
bins of width $\Delta f$, the optimal estimator and corresponding variance
are given by the weighted sum
\begin{align}
\hat\Omega_\alpha \equiv
\frac{\sum_f \sigma^{-2}_{\hat\Omega_\alpha}(f)\hat\Omega_\alpha(f)}
{\sum_{f'}\sigma^{-2}_{\hat\Omega_\alpha}(f')}
\,,
\quad
\sigma^{-2}_{\hat\Omega_\alpha} \equiv
\sum_f\sigma^{-2}_{\hat\Omega_\alpha}(f)\,.
\label{e:summed_estimator}
\end{align}
A similar weighted sum can be used to optimally combine the estimators
calculated for different time intervals \footnote{For ease of notation, we do
not explicitly show the time-dependence of the estimators.}.

In the presence of correlated noise, the estimators are biased. The expected
values are then
\begin{equation}
\langle \hat\Omega_\alpha(f)\rangle
=\Omega_\alpha+\eta_\alpha(f)\,,
\label{e:Omega_tot(f)}
\end{equation}
where
\begin{equation}
\eta_\alpha(f)
\equiv
\frac{\Re\left[N_{12}(f)\right]}{\gamma(f)S_\alpha(f)}\,.
\label{e:Omega_noise(f)}
\end{equation}
Here
$N_{12}(f)\equiv\frac{2}{T}\langle\tilde n_1^*(f)\tilde n_2(f)\rangle$
is the one-sided cross-spectral density (CSD) of the correlated noise
contribution $\tilde n_1, \tilde n_2$ to $\tilde s_1, \tilde s_2$.
The expression for the variance $\sigma_{\hat\Omega_\alpha}^2(f)$ is
unchanged in the presence of correlated noise provided
$|N_{12}(f)|\ll P_1(f), P_2(f)$. For the summed estimator
$\hat\Omega_\alpha$, we have
\begin{equation}
\langle \hat\Omega_\alpha\rangle
=\Omega_\alpha+\eta_\alpha
\label{e:Omega_tot}
\end{equation}
where
\begin{equation}
\eta_\alpha
\equiv
\frac{\sum_f \sigma^{-2}_{\hat\Omega_\alpha}(f)\eta_\alpha(f)}
{\sum_{f'}\sigma^{-2}_{\hat\Omega_\alpha}(f')}
\label{e:Omega_noise}
\end{equation}
is the contribution from correlated noise averaged over time (not shown) and
frequency. Thus, {\em correlated noise biases our estimates of the 
amplitude of a SGWB}. Here we also note that
$\eta_\alpha$ can be positive or negative while $\Omega_\alpha$ is positive
by definition. The purpose of the noise identification and removal methods
that we describe below 
is to reduce this bias as much as possible.

\section{Methods for identifying correlated noise}
\label{sec:methods}

\subsection{Coherence calculation}
\label{sec:coherence}

\begin{figure}
\includegraphics[width=3.35in]{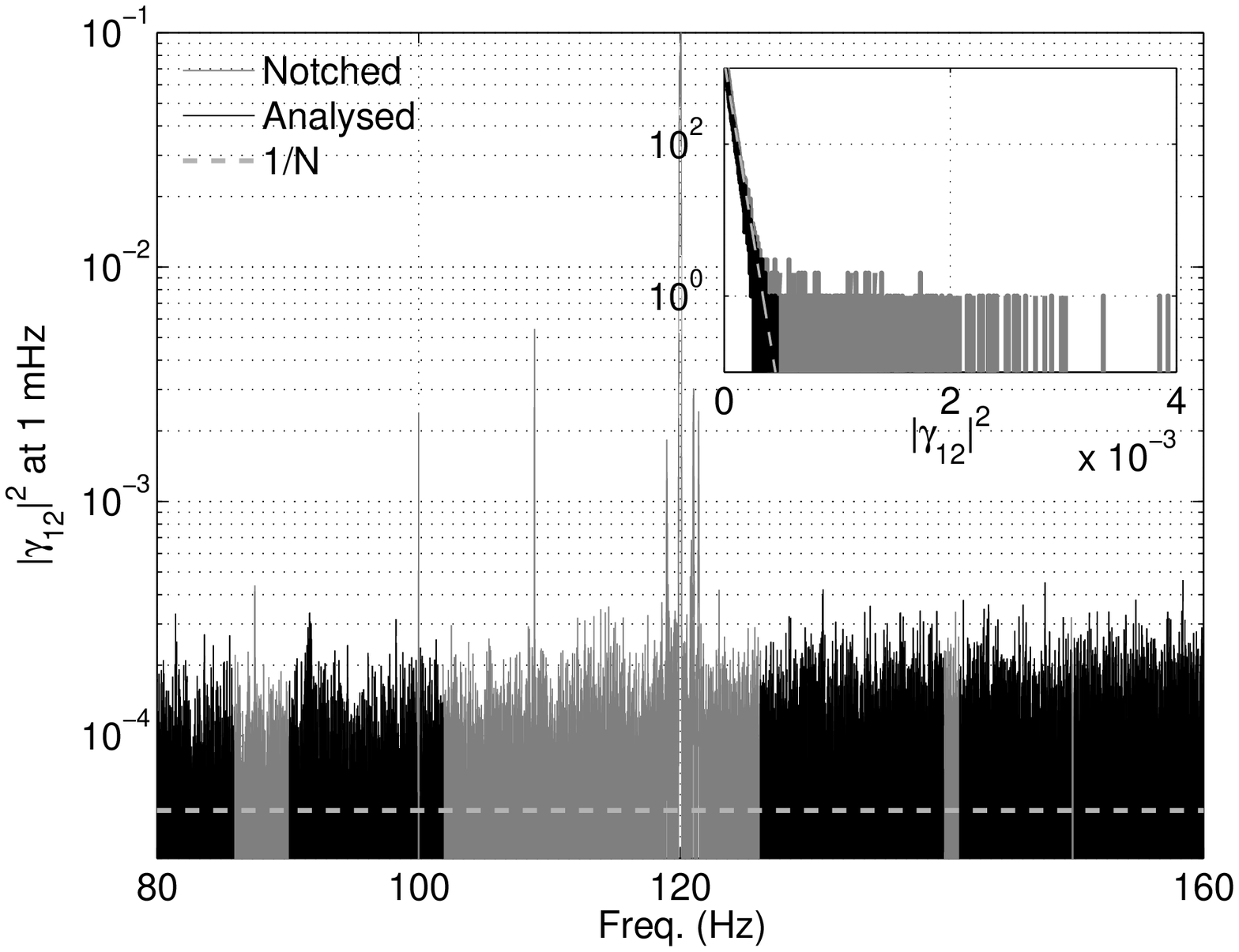}
\includegraphics[width=3.35in]{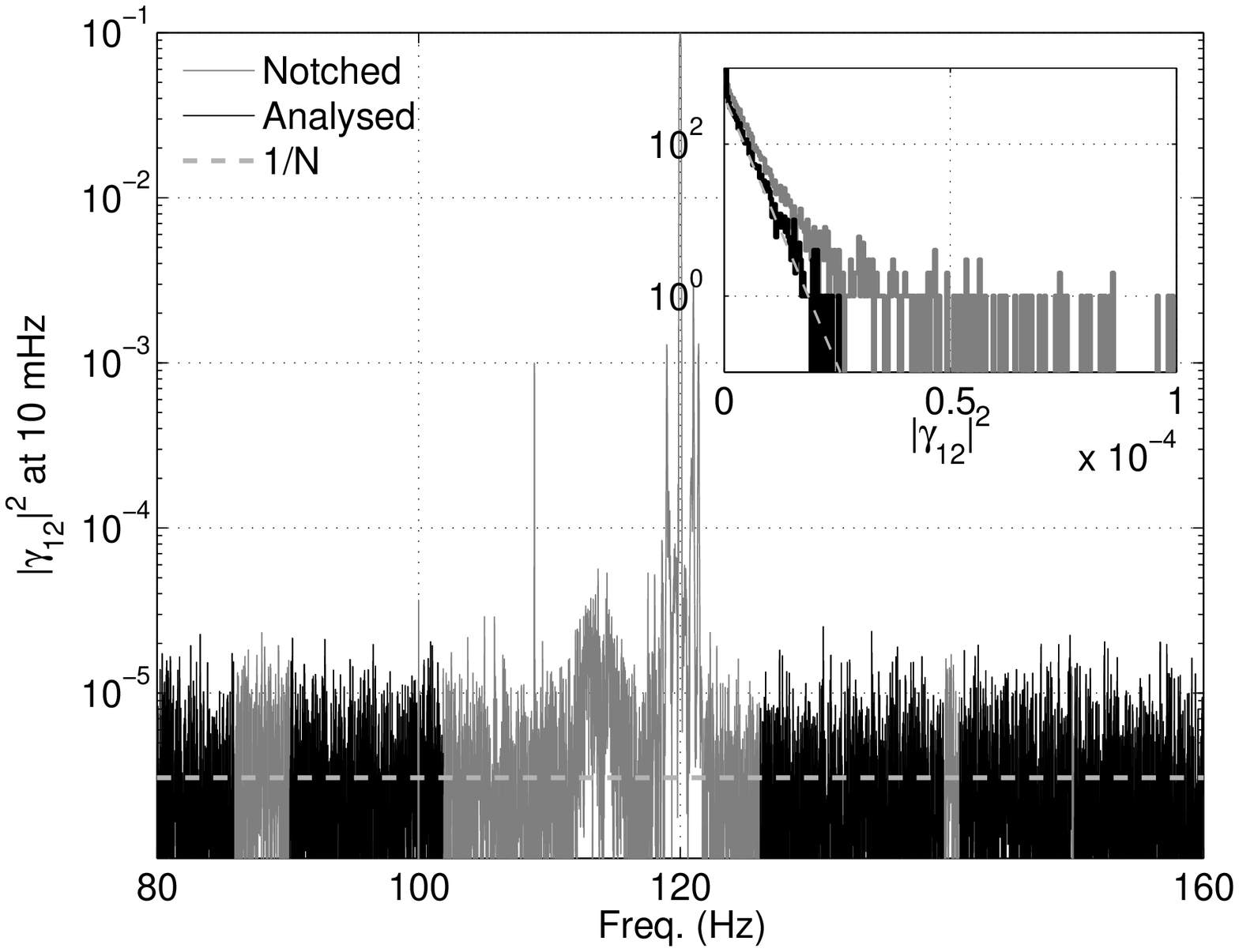}
\includegraphics[width=3.35in]{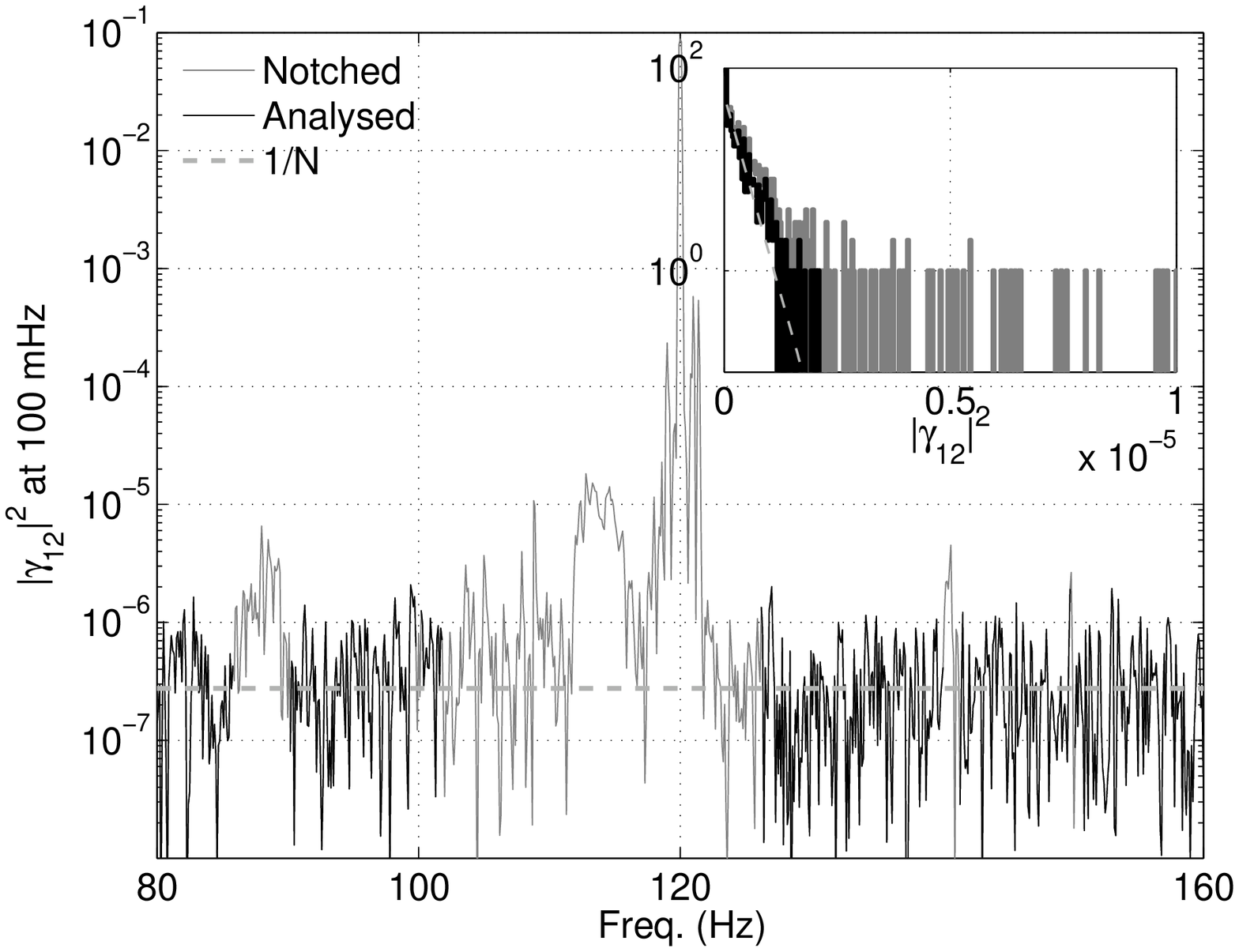}
\caption{Coherence $\hat{\Gamma}_{12}$ between H1 and H2 computed in the 
frequency band 80--160~Hz using all of the S5 data, for three different 
frequency resolutions: 1~mHz, 10~mHz, and 100~mHz (top-to-bottom). The 
insets show that the histograms of the coherence at the analyzed frequencies 
follow the expected exponential distribution for Gaussian noise, as well as 
the presence of a long tail of high coherence values at notched frequencies. A 
stochastic broadband GW signal of SNR = 5 would appear at a level of 
$\lesssim10\times$ below the dashed $1/N$ line.}
\label{f:coh_low}
\end{figure}
\begin{figure}
\includegraphics[width=3.35in]{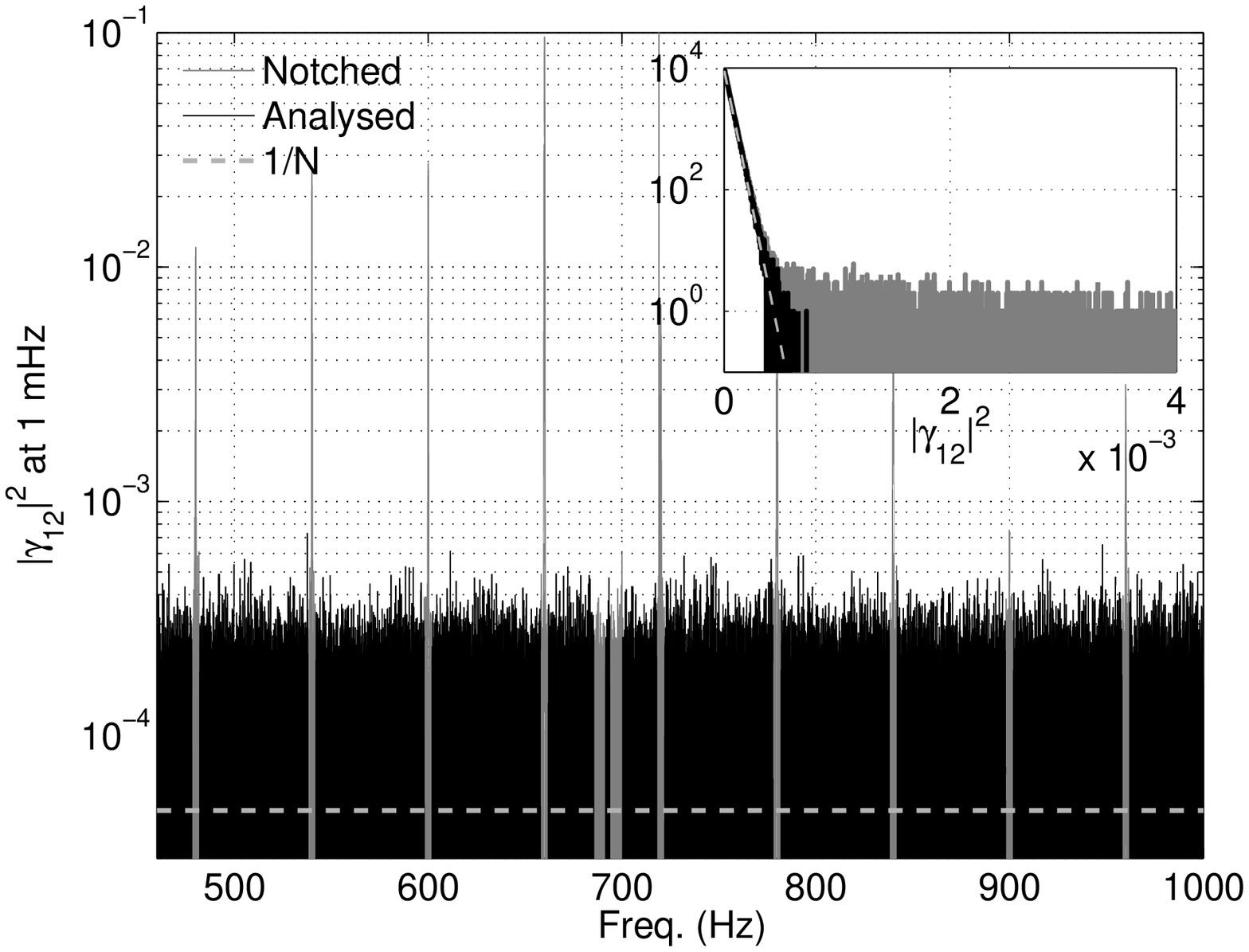}
\includegraphics[width=3.35in]{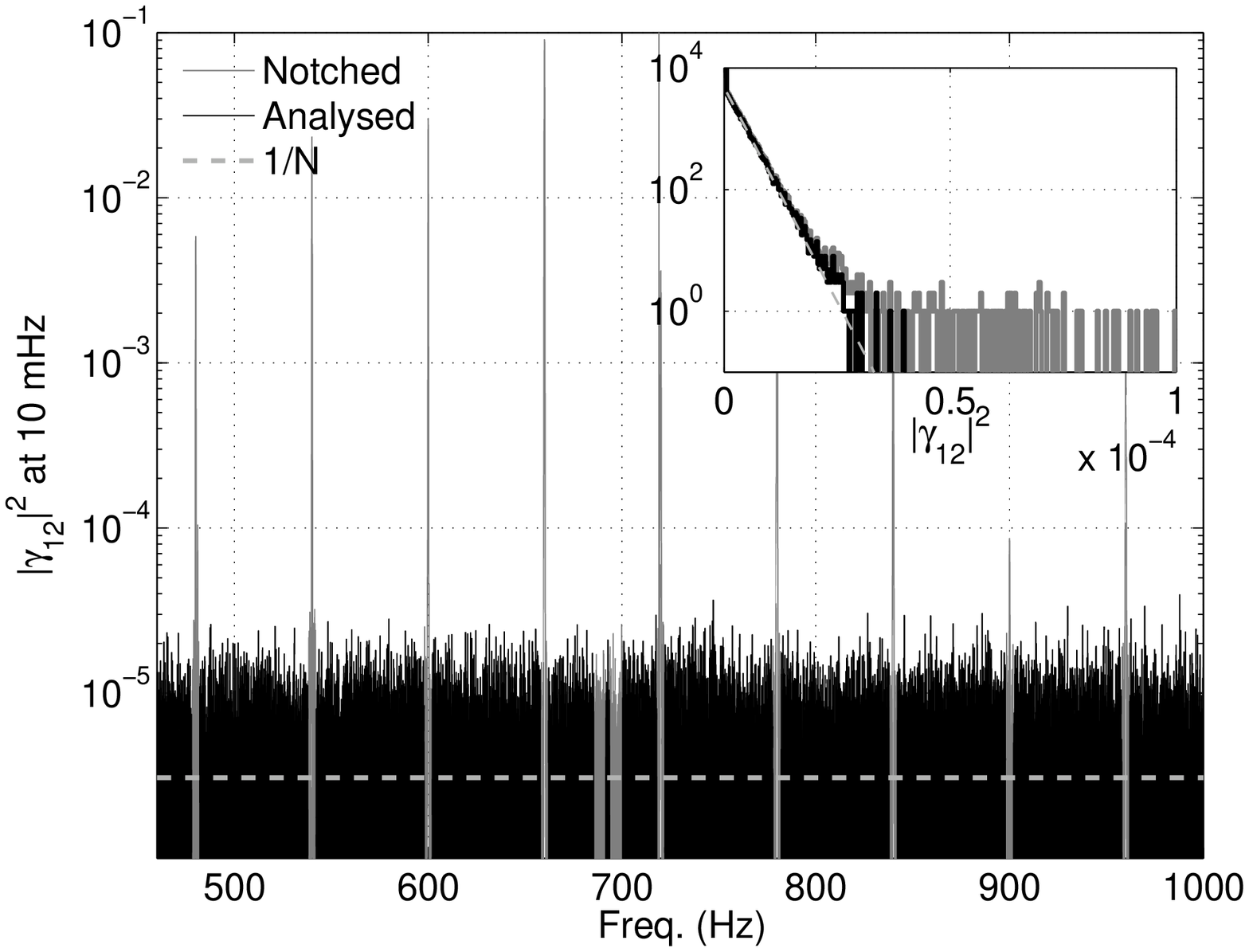}
\includegraphics[width=3.35in]{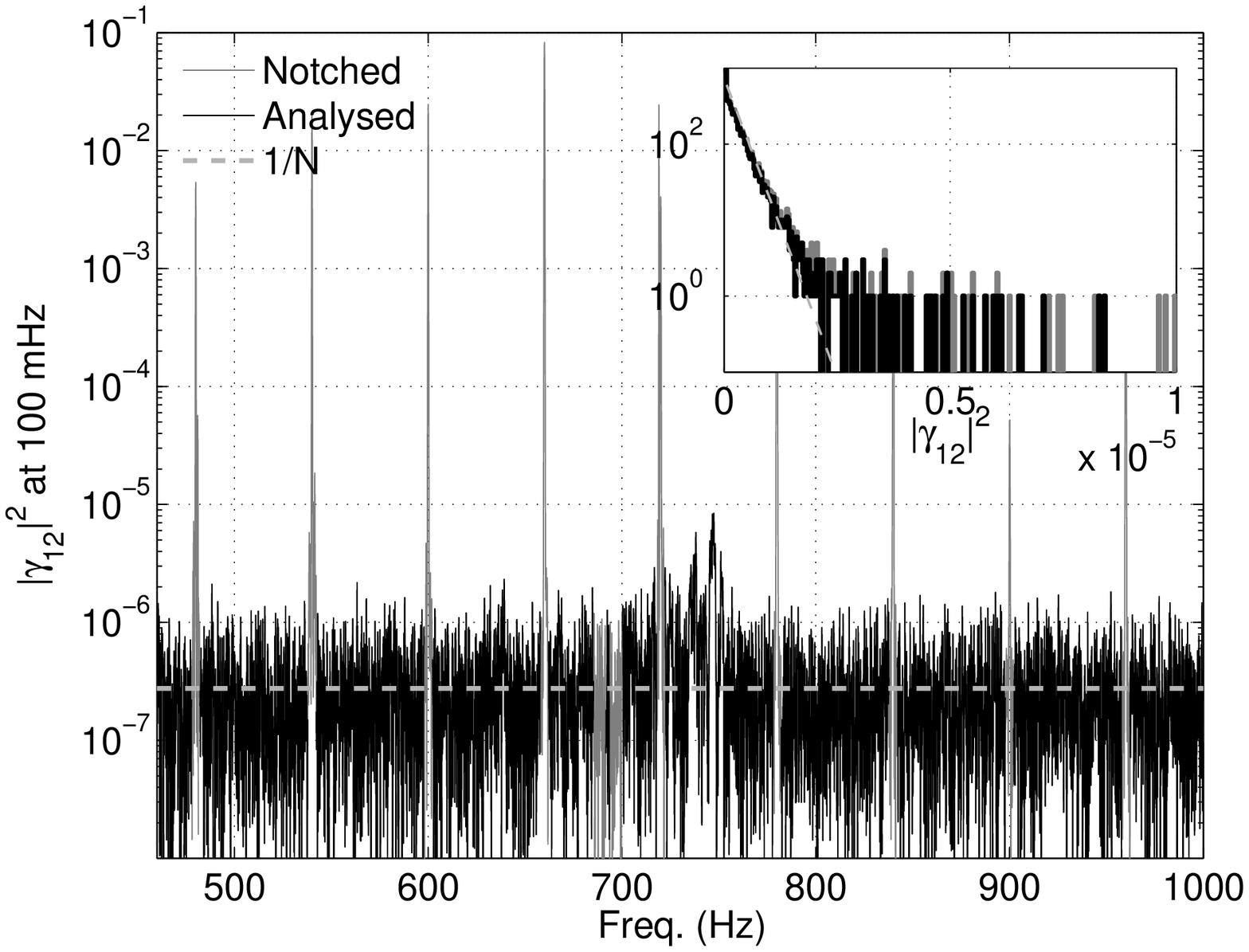}
\caption{Same as Fig.~\ref{f:coh_low} but at higher frequencies, 460--1000~Hz.
Note the coherence peaks at the harmonics of the 60~Hz power lines (notched
in the analysis). The elevated coherence near 750~Hz at 100~mHz resolution
is due to acoustic noise coupling to the GW channels. The long tail in the 
100 mHz plot is due to excess noise around 750~Hz, which was removed from the 
final analysis using PEM notchings (see Sec.~\ref{sec:steps}). A stochastic 
broadband GW signal of SNR = 5 would appear at a level of $\lesssim10\times$ 
below the dashed $1/N$ line.}
\label{f:coh_high}
\end{figure}

Perhaps the simplest method for identifying correlated noise in the H1-H2
data is to calculate the magnitude squared coherence,
$\hat{\Gamma}_{12}(f)\equiv |\gamma_{12}(f)|^2$,
where
\begin{equation}
\gamma_{12}(f)
\equiv
\frac{2}{T}\frac{\langle
\tilde s_1^*(f) \tilde s_2(f)\rangle_N}
{\sqrt{\langle P_1(f)\rangle_N
\langle P_2(f)\rangle_N}}\,.
\label{e:gamma12}
\end{equation}
Here $T$ denotes the duration of a single segment of data, and angle brackets
$\langle\ \rangle_N$ denotes an average over $N$ segments used
to estimate the CSD and PSDs that enter the expression for $\gamma_{12}$.
If there are no correlations (either due to noise or a GW a signal) in the
data, the expected value of $\hat{\Gamma}_{12}(f)$ is equal
to $1/N$. This method is especially useful at finding narrowband features
that stick out above the expected $1/N$ level. Since we expect a SGWB
to be broadband, with relatively little variation in the LIGO
band ($\sim$80--1000~Hz), most of these features can be attributed to
instrumental and/or environmental correlations. We further investigate these
lines with data from other PEM channels and once we confirm that they are
indeed environmental/instrumental artifacts, we remove them from our analysis.

Plots of $\hat{\Gamma}_{12}(f)$ for three different frequency resolutions
are shown in Figs.~\ref{f:coh_low} and \ref{f:coh_high} for two frequency
bands, 80--160~Hz and 460--1000~Hz, respectively.
In Fig.~\ref{f:coh_low}, note the relatively wide structure around 120~Hz,
which is especially prominent in the bottom panel where the frequency
resolution is 100~mHz. This structure arises from low-frequency noise
(dominated by seismic and other mechanical noise) up-converting to
frequencies around the 60~Hz harmonics via a bilinear coupling mechanism.
While these coupling mechanisms are not fully understood, we reject the band
from 102--126~Hz for our analysis, given the elevated correlated noise seen in
this band. (A similar plot at slightly lower and higher frequencies shows
similar noisy bands from 40--80~Hz and 160--200~Hz.) A closer look at the
coherence also identifies smaller structures at 86--90~Hz, 100~Hz,
140--141~Hz, and 150~Hz. A follow-up analysis of PEM channels (which is
discussed in more detail later) revealed that
the grayed bands in Figs.~\ref{f:coh_low} and \ref{f:coh_high} were
highly contaminated with acoustic noise or by low-frequency seismic noise
up-converting to frequencies around the 60~Hz harmonics via a bilinear
coupling mechanism; so we rejected these
frequency bands from subsequent analysis. As mentioned earlier, the
160--460 Hz band was not used in this analysis, because of similar acoustic
and seismic contamination, as well as violin-mode resonances of the
mirror-suspension wires (see Sec.~\ref{sec:comparePEMandTS}).

As shown in Fig.~\ref{f:coh_high}, the coherence at high frequencies
(460--1000~Hz) is relatively clean. The only evidence of narrow-band
correlated noise is in $\pm 2~{\rm Hz}$ bands around the 60-Hz power-line
harmonics, and violin-mode resonances of mirror suspensions at
$688.5\pm 2.8$ Hz and $697\pm 3.1$ Hz. The elevated coherence near
750~Hz at 100~mHz resolution is due to acoustic noise coupling to the GW
channels. Notching the power-line harmonics and violin-mode resonances
amounts to the removal of $\sim 9 \%$ of the frequency bins over the entire
high-frequency band.

\subsection{Time-shift analysis}
\label{sec:timeshift}

A second method for identifying narrowband correlated noise is to
{\em time-shift} the time-series output of one detector relative to that
of the other detector before doing the cross-correlation analysis
\cite{nicktime}. By introducing a shift of $\pm 1$~second, which is
significantly larger than the correlation time for a broadband GW signal
($\sim\!10~{\rm ms}$, cf.~Fig.~\ref{softinj}), we eliminate broadband GW
correlations while preserving narrowband noise features. Using segments of
duration $T=1~{\rm s}$, we calculate the time-shifted estimators
$\hat\Omega_{\alpha,{\rm TS}}(f)$, variance
$\sigma^2_{\Omega_\alpha, {\rm TS}}(f)$, and their ratio
${\rm SNR}_{\Omega_\alpha, {\rm TS}}(f)\equiv
\hat\Omega_{\alpha,{\rm TS}}(f)/\sigma_{\Omega_\alpha, {\rm TS}}(f)$.
The calibration and conditioning of the data is performed in exactly the same
way as for the final search, which is described in detail in
Secs.~\ref{sec:steps} and \ref{sec:results}.

We excise any frequency bin with $|{\rm SNR}_{\Omega_\alpha,{\rm TS}} (f)| >2$
on the grounds that it is likely contaminated by correlated noise. This
threshold was chosen on the basis of initial studies performed using
playground data to understand the
effectiveness of such cut. This criterion can be checked for different
time-scales, such as weeks, months, or the entire data set. This allows us to
identify transient effects on different time-scales, which may be diluted
(and unobservable) when averaged over the entire data set.

\subsection{PEM coherence calculations}
\label{sec:PEMcoherence}

Another method for identifying correlated noise is to first try to identify
the noise sources that couple into the individual detector outputs by
calculating the coherence of $\tilde s_1$ and $\tilde s_2$ with various
PEM channels $\tilde z_I$:
\begin{equation}
\hat{\gamma}_{iI}(f)\equiv
\frac{2}{T}
\frac{\langle \tilde s_i^*(f) \tilde z_I(f)\rangle_N}
{\sqrt{\langle P_i(f) \rangle_N\langle P_I(f)\rangle_N}}\,.
\end{equation}
Here $i=1,2$ labels the detector outputs and $I$ labels the PEM channels.
For our analysis we used 172 PEM channels located near the two
detectors. In addition to the PEM channels, we used a couple of auxiliary 
channels associated with the stabilization of the frequency of the lasers 
used in the detectors, which potentially carry information about instrumental 
correlations between the two detectors. (Hereafter, the usage of the acronym 
PEM will also include these two auxiliary channels.) The Fourier transforms
are calculated for each minute of data ($T=60~{\rm s}$), and the average CSDs
and PSDs are computed for extended time-periods---%
weeks, months, or the entire run. We then perform the following maximization
over all PEM channels, for each frequency bin $f$, defining:
\begin{equation}
\hat{\gamma}_{12,{\rm PEM}} (f)\equiv
\max_I \Re
\left[\hat{\gamma}_{1I}(f)\times\hat{\gamma}^*_{2I}(f)\right]\,.
\end{equation}
Note that by construction $\hat{\gamma}_{12,{\rm PEM}}(f)$ is real.

As discussed in \cite{nickcoh}, $\hat{\gamma}_{12,{\rm PEM}}(f)$ is an 
estimate of the instrumental or environmental contribution to the coherence 
between the GW channels of H1 and H2. This estimate is only approximate, 
however, and
potentially suffers from systematic errors for a few reasons. First, the
PEM coverage of the observatory may be incomplete---i.e., there may be
environmental or instrumental effects that are not captured by the existing
array of PEMs. Second, some of the PEM channels may be correlated.
Hence, a rigorous approach would require calculating a matrix of elements
$\hat{\gamma}_{IJ}(f)$, and then inverting this matrix or solving a set of 
linear equations involving elements of $\hat{\gamma}_{IJ}(f)$.
In practice, due to the large number of
channels and the large amount of data, this is a formidable task. Instead, we
simply maximize, frequency-by-frequency, over the contributions from different
PEM channels and use this maximum as an estimate of the overall environmental
contribution to $\hat{\gamma}_{12}(f)$. Finally, these coherence methods 
do not take into account the nonlinear upconversion processes in which 
low-frequency disturbances, primarily seismic activity, excite 
higher-frequency modes in the instrument.

Since the measured signal-to-noise ratio for the estimator
$\hat\Omega_\alpha(f)$ can be written as
\begin{align}
{\rm SNR}(f)
=\sqrt{2T\Delta f}\,
\Re\left[\hat{\gamma}_{12}(f)\right]\,,
\end{align}
we can simply approximate the contribution of the PEM channels to the
stochastic GW signal-to-noise ratio as
\begin{equation}
{\rm SNR}_{\rm PEM}(f)
\equiv
\sqrt{2T\Delta f}\,
\hat{\gamma}_{12,{\rm PEM}}(f)\,,
\end{equation}
remembering that $\hat{\gamma}_{12,{\rm PEM}}(f)$ is real. The PEM 
contribution to the estimators $\hat\Omega_\alpha(f)$ is then
\begin{equation}
\hat\Omega_{\alpha,{\rm PEM}}(f)
\equiv
{\rm SNR}_{\rm PEM}(f)
\sigma_{\hat\Omega_\alpha}(f)
\end{equation}
where $\sigma_{\hat\Omega_\alpha}(f)$ is the statistical uncertainty defined by
Eq.~\ref{e:var_Omega_est}.

We can use the PEM coherence calculations in two complementary ways. First,
we can identify frequency bins with particularly large instrumental or
environmental contributions by placing a threshold on
$|\mathrm{SNR}_{\rm PEM} (f)|$ and exclude them from the analysis.
Second, the frequency bins that pass this data-quality cut may still contain
some residual environmental contamination. We can estimate at least part of
this residual contamination by using $\hat\Omega_{\alpha,{\rm PEM}}(f)$ for
the remaining frequency bins.

As part of the analysis procedure, we were able to identify the PEM channels
that were responsible for the largest coherent noise between the GW channels
in H1 and H2 for each frequency bin. For both the low and high frequency
analyses, microphones and accelerometers in the central building near the
beam splitters of each interferometer registered the most significant noise.
Within approximately 1~Hz of the 60-Hz harmonics, magnetometers and voltage
line monitors registered the largest correlated noise, but these frequencies
were already removed from the analysis due to the significant coherence
(noise) level at these frequencies, as mentioned in Sec.~\ref{sec:coherence}.

\subsection{Comparing PEM-coherence and time-shift methods}
\label{sec:comparePEMandTS}

\begin{figure}
\includegraphics[width=3.35in]{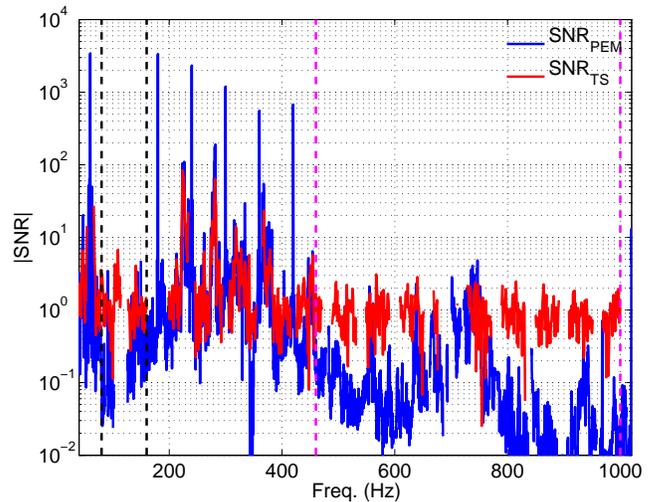}
\caption{(Color online) Comparison of the (absolute value of the) SNRs 
calculated by the
PEM-coherence and the time-shift techniques. The vertical dotted lines
indicate the frequency bands used for the low (80--160 Hz; black dotted lines)
and high (460--1000 Hz; magenta dotted lines) frequency analyses. Note that
$\mathrm{SNR}_{\Omega_\alpha,{\rm TS}}(f)$ is a true signal-to-noise ratio,
so values $\lesssim 2$ are dominated by random statistical fluctuations.
$\mathrm{SNR}_{\Omega_\alpha,{\rm PEM}}(f)$, on the other hand, is an estimate
of the PEM contribution to the signal-to-noise ratio, so values even much
lower than 2 are meaningful measurements (i.e., they are not statistical
fluctuations). The two methods agree very well in
identifying contaminated frequency bins or bands. Note that both methods
indicate that the 80--160~Hz and 460--1000~Hz bands have relatively
low levels of contamination.
}
\label{f:TSPEM}
\end{figure}

Figure~\ref{f:TSPEM} shows a comparison of the SNRs calculated by the
PEM-coherence and time-shift methods. The agreement between these two very
different techniques in identifying contaminated frequency bins (those
with $|SNR| \gtrsim$ a few) is
remarkably good, which is an indication of their robustness and effectiveness.
Moreover, Fig.~\ref{f:TSPEM} shows that the frequency region between 200~Hz
and 460~Hz is particularly contaminated by environmental and/or instrumental
effects. Hence, in this analysis we focus on the low-frequency region
(80--160~Hz) which is the most sensitive to cosmological backgrounds (i.e.,
spectral index $\alpha=0$), and on the high-frequency region (460--1000~Hz)
which is less contaminated and more suitable for searches for
astrophysically-generated backgrounds (e.g., $\alpha=3$).

We emphasize that the PEM channels only monitor the instrument and the
environment, and are not sensitive to GWs. Similarly, the time-shift analysis,
with a time-shift of $\pm 1$ second, is insensitive to broad-band GW signals.
Hence, any data-quality cuts based on the PEM and time-shift studies will
not affect the astrophysical signatures in the data---i.e., they do not bias
our estimates of the amplitude of a SGWB.

\subsection{Other potential non-astrophysical sources of correlation}
\label{sec:othersources}

We note that any correlations that are produced by environmental signals that 
are not detected by the PEM sensors will not be detected by the 
PEM-coherence technique. Furthermore, if such correlations, or correlations 
from a non-environmental source, are broadband and flat (i.e., do not vary 
with frequency over our band), they will not be detected by either the 
PEM-coherence or the time-shift method. One potential source of broadband
correlation between the two GW channels is the data acquisition system itself.
We investigated this possibility by looking for correlations
between 153 channel pairs that had no physical reason to be correlated.
We found no broadband correlations, although we did find an
unexplained narrow-band correlation at 281.5~Hz between 10 of
153 channel pairs. Note that 281.5 Hz is outside of the frequency bands
analyzed in this study.

We addressed the potential of correlations from un-monitored environmental 
signals by searching for coupling sites four times over the course of the 
run by injecting large but localized acoustic, seismic, magnetic and RF 
signals. New sensors were installed at the two coupling sites that had
the least coverage. However, we found that the new sensors, even after 
scaling up to the full analysis period, contribute less than 1\% of the 
total frequency notches; hence it is safe to assume that we had sufficient PEM
coverage throughout our analysis period.

We also examined the possibility of correlations between the H1 and H2 
detectors being generated by scattered light. We considered two mechanisms:
first, light scattered from one detector affecting the other detector, and 
second, light from both detectors scattering off of the same site and 
returning to the originating detectors. We did not observe, and do not expect 
to observe, the first mechanism because the frequencies of the two lasers,
while very stable, may differ by gigahertz. If light from one interferometer 
scatters into the main beam of the other, it will likely be at a very different
frequency and will not produce signals in our 8 kHz band when it beats against 
the reference light for that interferometer.

Nevertheless, we checked for a correlation produced by light from one detector 
entering the other by looking for the calibration 
signals~\cite{LIGOdescription} injected into one detector in the
signal of the second detector. During S5, the following calibration line 
frequencies were injected into H1 and H2: 46.70~Hz, 393.10~Hz, 1144.30~Hz (H1) 
and 54.10~Hz, 407.30~Hz, 1159.7~Hz (H2). We note here that all those 
frequencies are outside of our analysis bands. We observed no correlation 
beyond the statistical error of the measurement at any of the three 
calibration line frequencies for either of the two detectors. This check was 
done for every week and month and for the entire S5 data-set. Hence, we 
conclude that potential signals carried by the light in one detector are not 
coupled into the other detector.

In contrast, we have observed the second scattering mechanism, in which 
scattered light from the H1 beam returns to the H1 main beam and H2 light 
returns to the H2 main beam. This type of scattering can produce H1-H2 
coupling if scattered light from H1 and from H2 both reflect off of the same 
vibrating surface (which modulates the length of the scattering paths) before
recombining with their original main beams. This mechanism is thought to 
account for the observation that shaking the reflective end cap of the 4 km 
beam tube (just beyond an H1 end test mass), produced a shaking-frequency 
peak in both H1 and H2 GW channels, even though the nearest H2 component was 
2 km away. However, this scattering mechanism is covered by the PEM system 
since the vibrations that modulate the beam path originate in the monitored 
environment.

We tested our expectation that scattering-induced correlations would be 
identified by our PEM-coherence method. We initiated a program to identify 
the most important scattering sites by mounting shakers on the vacuum 
system at 21 different locations that were selected as potential scattering 
sites, and searching for the shaking signal in the GW channels. All 
significant scattering sites that we found in this way were well-monitored by 
the PEM system. At the site that produced the greatest coherence between the
two detectors (a reflective flange close to and perpendicular to the beam 
paths of both interferometers), we mounted an accelerometer and found that 
the coherence between this accelerometer and the two GW channels was no greater
than that for the sensors in the pre-existing sensor system. These results 
suggest that the PEM system adequately monitored scattering coupling. As we 
shall show in Sec.~\ref{sec:highfrequency} below, no correlated noise (either
environmental or instrumental, either narrow-band or broad-band)
that is not adequately covered by the PEM system is identified in
the high-frequency analysis, further solidifying the adequacy of PEM system.

\section{Analysis procedure}
\label{sec:steps}

In the previous section we described a number of methods for identifying
correlated noise when searching for a SGWB. Here we
enumerate the steps for selecting the time segments and frequency bands
that were subsequently used for the analysis.

\smallskip
{\bf STEP 1}:
We begin by selecting time periods that pass a number of data quality flags.
In particular, we reject periods when:
(i) there are problems with the calibration of the data;
(ii) the interferometers are within 30~s of loss of servo control;
(iii) there are artificial signals inserted into the data for
calibration and characterization purposes;
(iv) there are PEM noise injections;
(v) various data acquisition overflows are observed; or
(vi) there is missing data.
With these cuts, the intersection of the H1 and H2 analyzable time
was $\sim 462$ days for the S5 run.

\smallskip
{\bf STEP 2:}
After selecting suitable data segments, we make a first pass at determining
the frequency bins to use in the analysis by calculating the overall
coherence between the detector outputs as described in Sec.~\ref{sec:coherence}.
Excess coherence levels led us to reject the frequencies
86--90~Hz, 100~Hz, 102--126~Hz, 140.25--141.25~Hz, and 150~Hz
in the low-frequency band (80--160 Hz), as well as
$\pm 2~{\rm Hz}$ around the 60~Hz power-line harmonics and the violin-mode
resonances at
$688.5\pm 2.8~{\rm Hz}$ and $697\pm 3.1~{\rm Hz}$ in the high-frequency
band (460--1000 Hz). It also identified a period of about 17 days in
June 2007 (between GPS times 866526322 and 867670285), during which the
detector H2 suffered from excessive transient noise glitches. We reject that
period from the analysis.

\smallskip
{\bf STEP 3:}
We perform a search for transient excess power in the data using the
wavelet-based Kleine Welle algorithm \cite{KW}, which was originally designed
for detecting GW bursts. This algorithm is applied to the output of both
detectors, producing a list of triggers for each detector. We then search the
two trigger lists and reject any segment that contains transients with
Kleine Welle significance larger than 50 in either of the two detectors. The
value of 50 is a conservative threshold, chosen based on other studies done on
the distribution of
such triggers in S5~\cite{S5glitch}.

\smallskip
{\bf STEP 4:}
Having determined the reasonably good frequency bands, we then calculate
$\hat\Omega_{\alpha}$ and its uncertainty $\sigma_{\hat\Omega_\alpha}$
summed over the whole band, cf.\ Eq.~\ref{e:summed_estimator}.
The purpose of this calculation is to perform another level of data-quality
selection in the time-domain by identifying noisy segments of 60~s duration. It
is similar to the non-stationarity cut used in the previous analyses
\cite{S1HLiso, S4HLiso, S5HLiso, S5-VSR1-LIGO-Virgo} where we remove time
segments whose $\sigma_{\hat\Omega_\alpha}$ differs, by a pre-determined
amount, from that calculated by averaging over two neighboring segments.
Here we use a $20\%$
threshold on the difference. The combination of the time-domain data quality
cuts described in Steps 1--4 removed about 22\% of the available S5
H1-H2 data.

\smallskip
{\bf STEP 5:}
After identifying and rejecting noisy time segments and frequency bins using
Steps 1--4, we then use the time-shift and the PEM-coherence methods
described in Secs.~\ref{sec:timeshift} and \ref{sec:PEMcoherence}
to identify any remaining contaminated frequency bins. To remove bad frequency
bins, we split the S5 dataset
into week-long periods and for each week, we reject any frequency bin for which
either $|{\rm SNR}_{\Omega_\alpha,{\rm TS}}(f)|$ or
$|{\rm SNR}_{\Omega_\alpha,{\rm PEM}}(f)|$ exceeds a pre-determined threshold
in the given week, the corresponding month, or in the entire S5 dataset.
This procedure generates (different) sets of frequency notchings for each week
of the S5 dataset. In the analysis we use two different sets of SNR threshold
values for the cut, which are further described in Sec.~\ref{sec:results}.

\begin{figure*}
\begin{tabular}{cc}
\includegraphics[angle=0,width=3.2in]{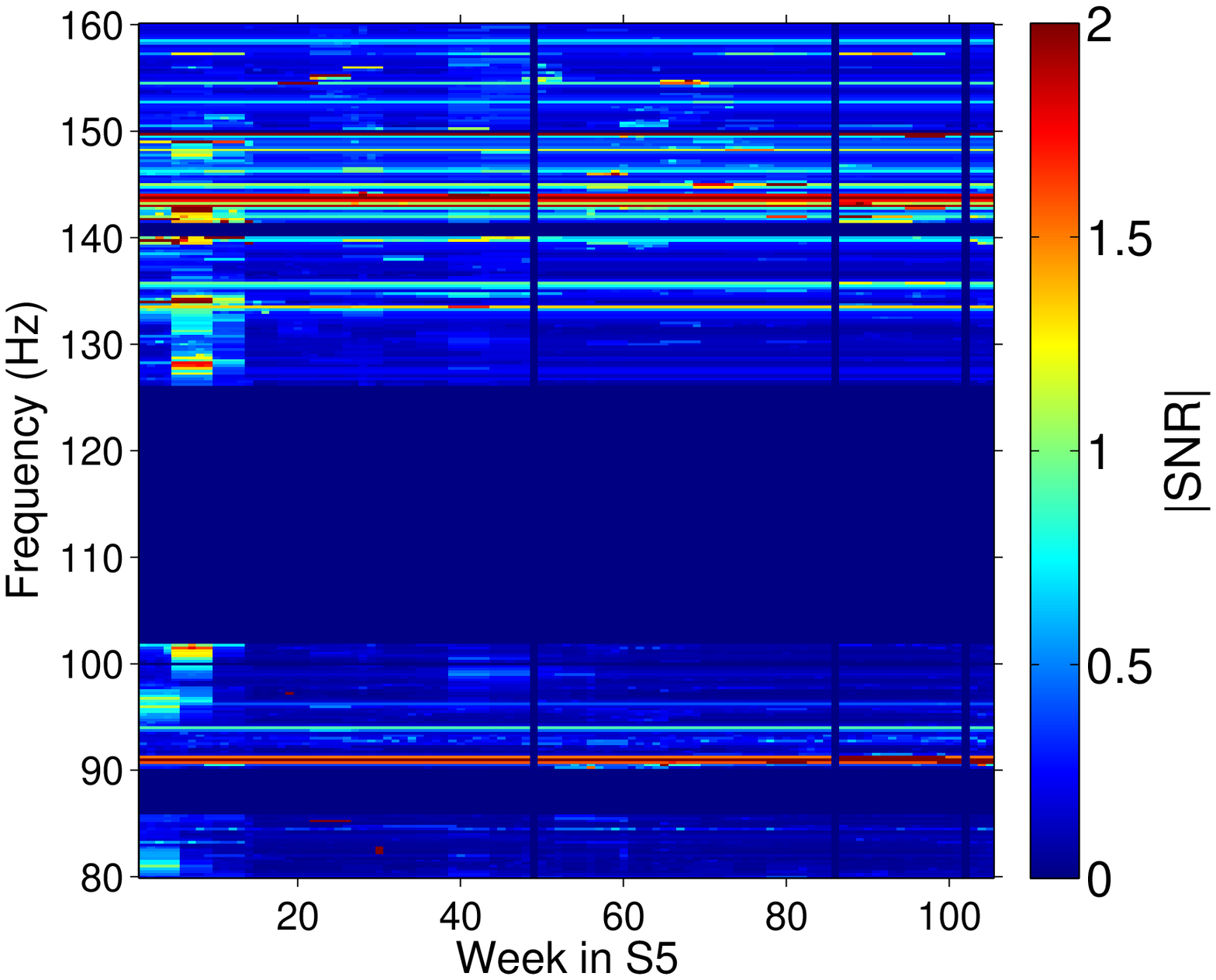}
\includegraphics[angle=0,width=3.2in]{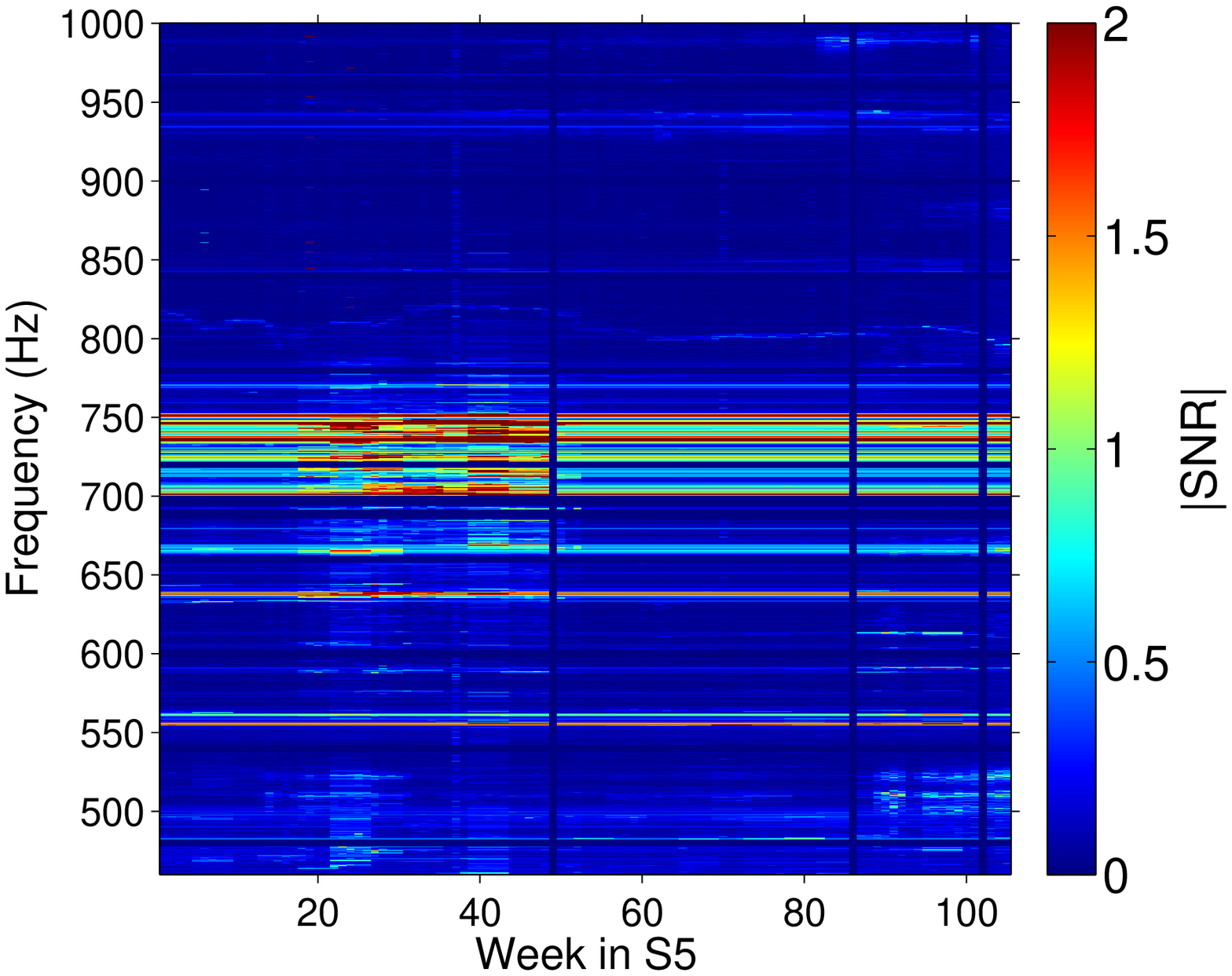}
\end{tabular}
\caption{(Color online) Spectrograms displaying the absolute value of
${\rm SNR}_{\Omega,{\rm PEM}}(f)$ for 80--160~Hz (left) and 460--1000~Hz
(right) as a function of the week in S5. The horizontal dark (blue) bands
correspond to
initial frequency notches as described in STEP 2 (Sec.~\ref{sec:steps}) and
vertical dark (blue) lines correspond to unavailability of data due to detector
downtime. The large SNR structures seen in the plots were removed from
the low- and high-frequency analyses.}
\label{PEM-contour}
\end{figure*}

Figure~\ref{PEM-contour} is a spectrogram of
${\rm SNR}_{\Omega_0,{\rm PEM}}$ for the 80--160~Hz band for all weeks in S5;
the visible structure represents correlated noise between H1 and H2, which was
identified and subsequently excluded from the analysis by the H1-H2 coherence,
time-shift, and PEM-coherence measurements.

Note that previous stochastic analyses using LIGO data
\cite{S1HLiso, S4HLiso, S5HLiso, S5-VSR1-LIGO-Virgo} followed only steps 1,
2 and 4.
Steps 3 and 5 were developed for this particular analysis.

\smallskip
Having defined the time-segments and frequency-bins to be rejected in each
week of the S5 data, we proceed with the calculation of the estimators and
standard errors, $\hat\Omega_\alpha(f)$ and $\sigma_{\hat\Omega_\alpha}(f)$,
in much the same manner as in previous searches for isotropic stochastic
backgrounds \cite{S5HLiso,S4HLiso,S3HLiso,S1HLiso}. The data is divided into
$T=60~{\rm s}$ segments, decimated to 1024~Hz for the low-frequency analysis
and 4096 Hz for the high-frequency analysis, and high-pass filtered with
a $6^{\rm th}$ order Butterworth filter with 32~Hz knee frequency. Each
analysis segment is Hann-windowed, and to recover the loss of signal-to-noise
ratio due to Hann-windowing, segments are 50\% overlapped. Estimators and standard
errors for each
segment are evaluated with a $\Delta f=0.25~{\rm Hz}$ frequency resolution,
using the frequency mask of the week to which the segment belongs.
A weighted average is performed over all segments and all frequency bins,
with inverse variances, as in Eq.~\ref{e:summed_estimator}, but properly
accounting for overlapping.

\section{Analysis results}
\label{sec:results}

The analysis is separated into two parts corresponding to searches for
SGWBs with spectral index $\alpha = 0$ and $\alpha = 3$ as described in
Sec.~\ref{sec:cross-correlation}. Since the strain output of an interferometer
due to GWs is $S_{\rm gw}(f) \propto f^{\alpha-3}$ (see Eq.~\ref{e:Sgw}),
the case $\alpha=0$ is dominated by low frequencies while $\alpha=3$
is independent of frequency. Since for $\alpha=3$ there is no preferred 
frequency band, and since previous analyses~\cite{S5-VSR1-LIGO-Virgo} for 
stochastic backgrounds with $\alpha=3$ considered only high frequencies, we 
also used only high frequencies for the $\alpha=3$ case. Thus, the two cases of
$\alpha=0$ and $\alpha=3$ correspond to the analysis of the low and
high-frequency bands, respectively. In this section, we present the results 
of the analyses in the two different frequency bands as defined in 
Sec.~\ref{sec:comparePEMandTS} corresponding to the two different values 
of $\alpha$.

To illustrate the effect of the various noise removal methods described in
the previous two sections, we give the results as different stages of cuts
are applied to the data (see Table~\ref{t:stages}).
The threshold value used at stage III comes from an initial study performed
using playground data to understand the effectiveness of the
PEM-coherence method in finding problematic frequency bins in the H1-H2 
analysis, and hence those
results are considered as {\it blind analysis} results. But a post-unblinding
study showed that we could lower the ${\rm SNR}_{\rm PEM}$ threshold to values
as low as 0.5 (for low-frequency) and 1 (for high-frequency), which are used at
stage IV. These post-blinding results are used in the final upper-limit
calculations. For threshold values $<0.5$ (low-frequency) or $<1$
(high-frequency), the PEM-coherence contribution,
$\hat\Omega_{\alpha,{\rm PEM}}$, varies randomly as the threshold is changed
indicating the statistical noise limit of the PEM-coherence method.
\begin{table}[h!]
\centering
\begin{tabular}{|c|l|c|l|c|}
\hline
\multirow{3}{*}{Stage} & \multicolumn{2}{|c|}{High-frequency analysis} & \multicolumn{2}{|c|}{Low-frequency analysis} \\
\cline{2-5}
 & \multirow{2}{*}{Steps} & \% of data & \multirow{2}{*}{Steps} & \% of data \\
 &       & vetoed & & vetoed \\
\hline
I & Step 1 & 8.51 & Step 1 & 8.51\\
\hline
II & Steps 1--4 & 35.88 & Steps 1--4 & 56.01 \\
\hline
III & Steps 1--5 with & 47.19 & Steps 1--5 with & 72.29\\
& $|{\rm SNR}_{\rm PEM}|>2$, & & $|{\rm SNR}_{\rm PEM}|>2$,& \\
& $|{\rm SNR}_{\rm TS}|>2$ & & $|{\rm SNR}_{\rm TS}|>2$ &\\
\hline
IV & Steps 1--5, with & 48.95 & Steps 1--5, with & 76.60\\
& $|{\rm SNR}_{\rm PEM}|>1$, & & $|{\rm SNR}_{\rm PEM}|>0.5$, & \\
& $|{\rm SNR}_{\rm TS}|>2$ & & $|{\rm SNR}_{\rm TS}|>2$ &\\
\hline
\end{tabular}
\caption{Definition of various stages of noise removal for the high and
low-frequency analyses in terms of the analysis steps described in
Sec.~\ref{sec:steps}. Here stage III corresponds to the blind analysis
and stage IV to the post-unblinding analysis. The percentage
of data vetoed accounts for both the time segments and frequency bins excluded
from the analysis. In calculating veto percentage, the analyses with
non-colocated LIGO detectors
only accounts for the time segments excluded from the analyses and is the
reason for the large numbers we see in the last column compared to other LIGO
analyses.}
\label{t:stages}
\end{table}

\subsection{High-frequency results}
\label{sec:highfrequency}

\begin{figure*}
\begin{tabular}{cc}
\includegraphics[angle=0,width=2.6in]{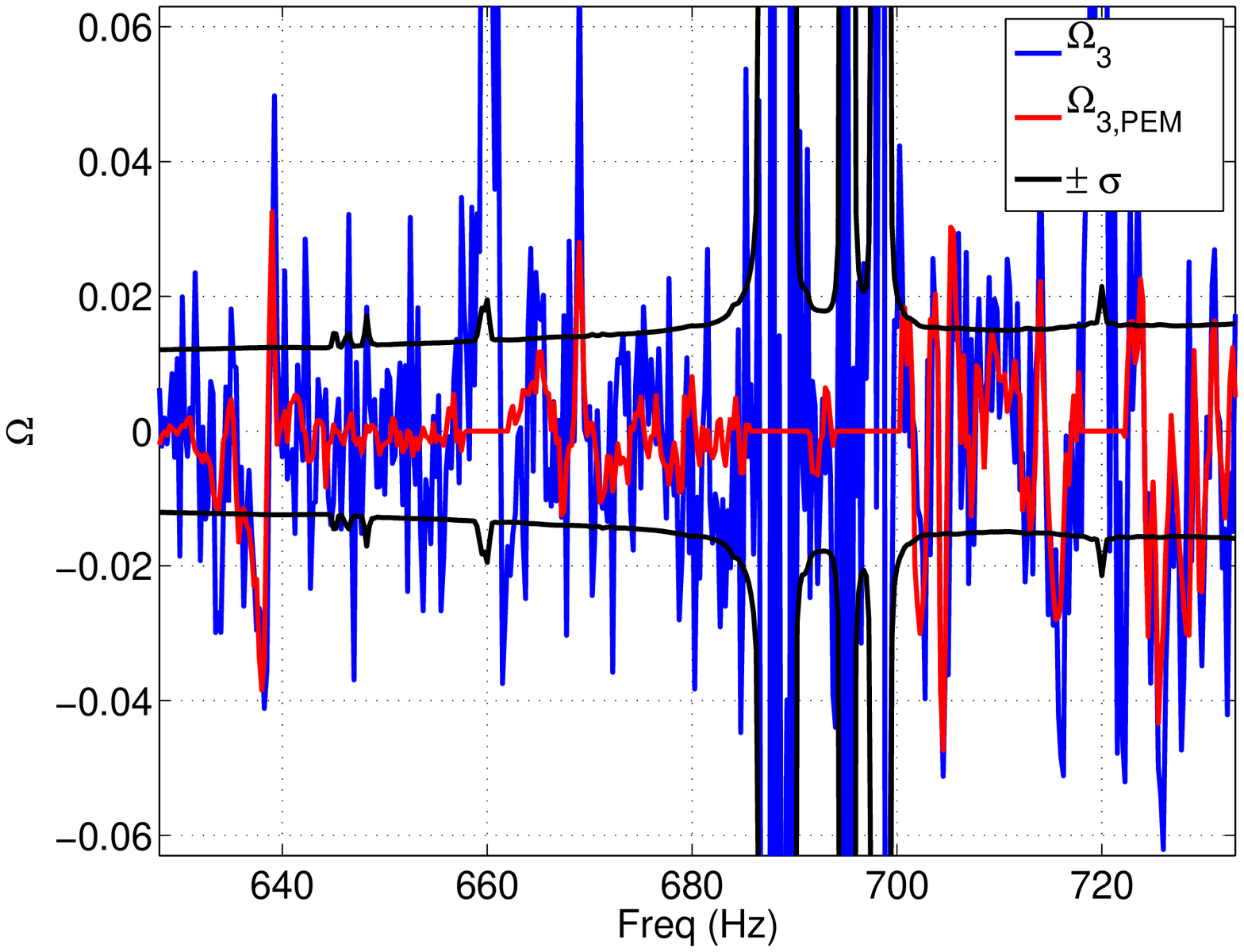}
\includegraphics[angle=0,width=2.6in]{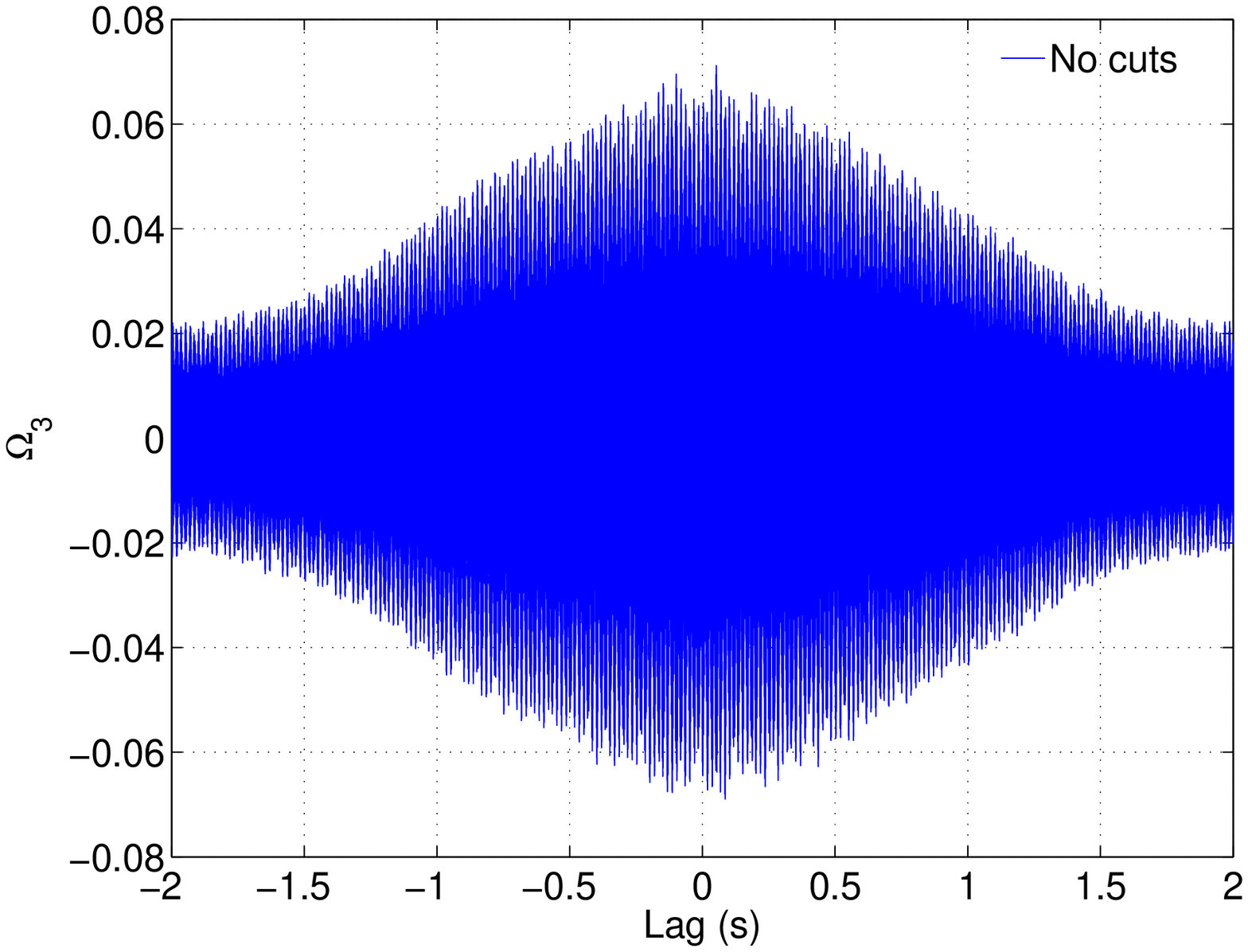}
\\
\includegraphics[angle=0,width=2.6in]{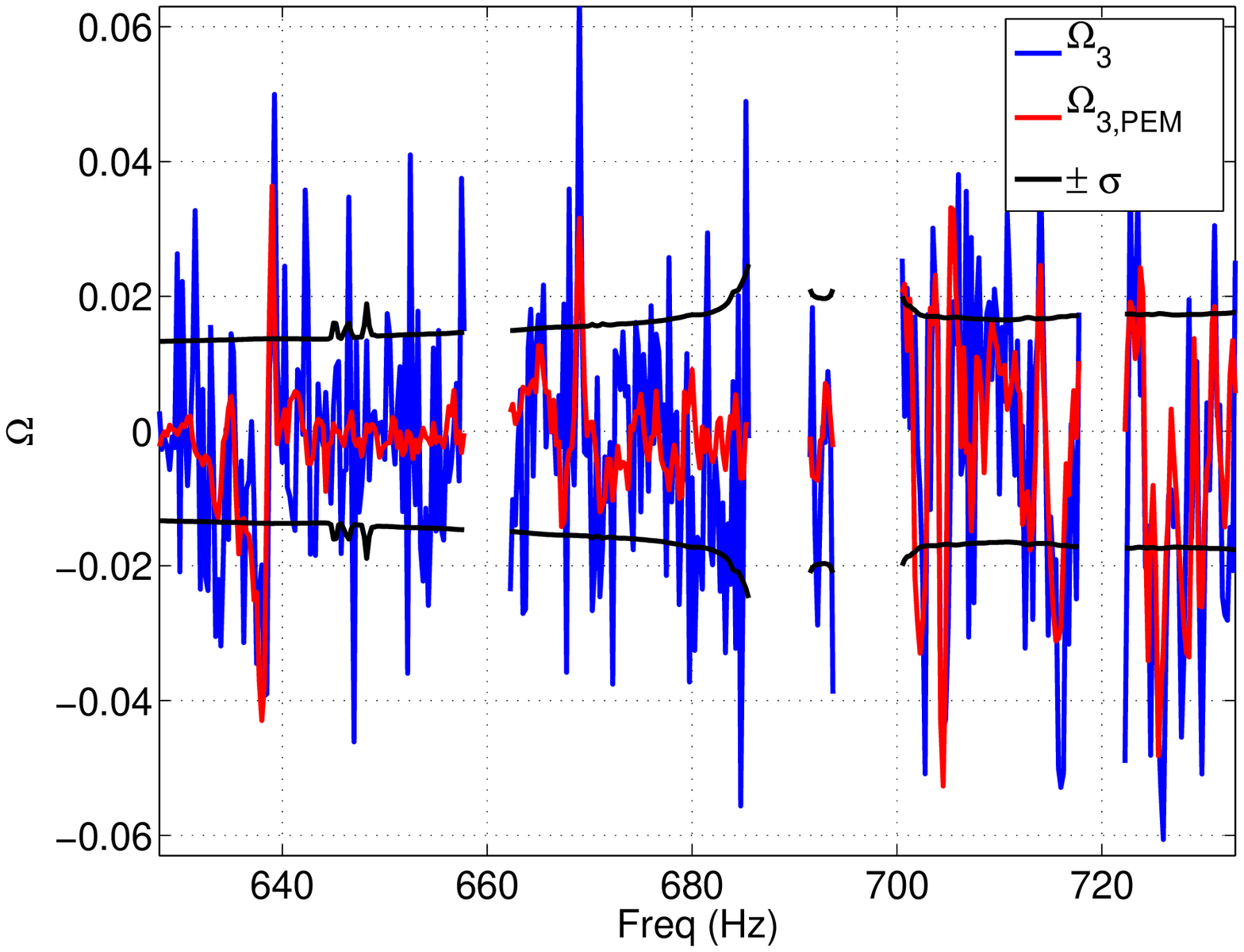}
\includegraphics[angle=0,width=2.6in]{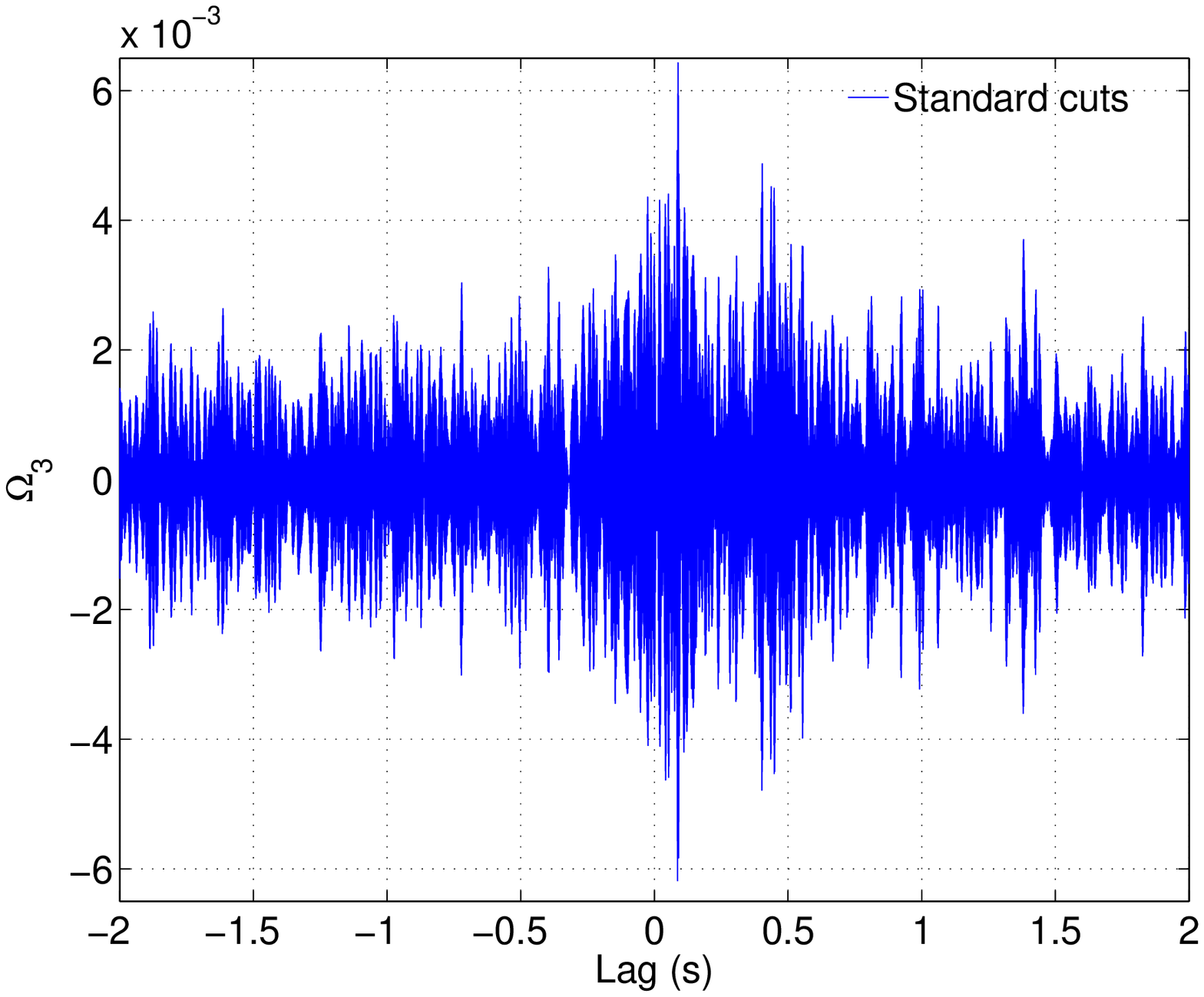}
\\
\includegraphics[angle=0,width=2.6in]{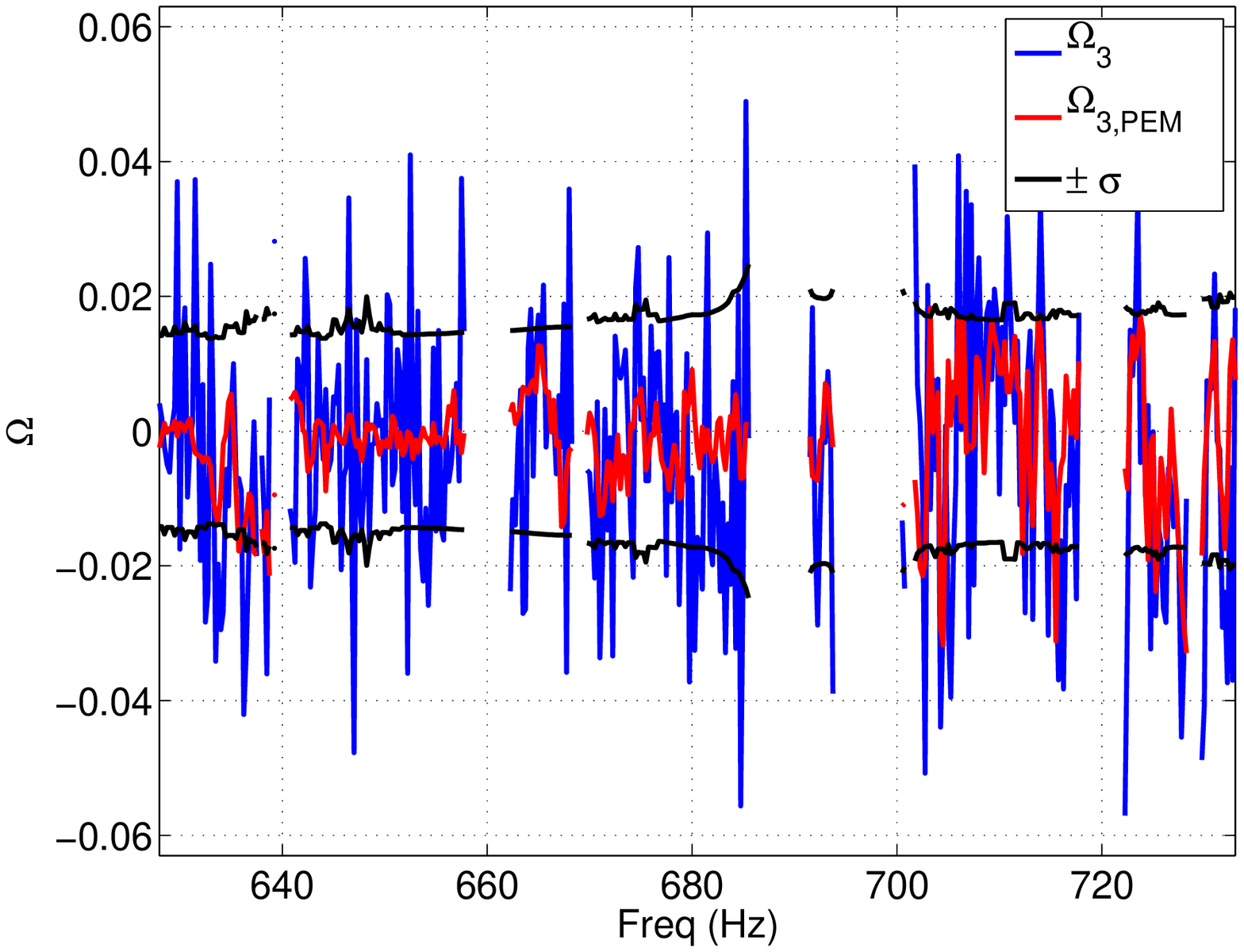}
\includegraphics[angle=0,width=2.6in]{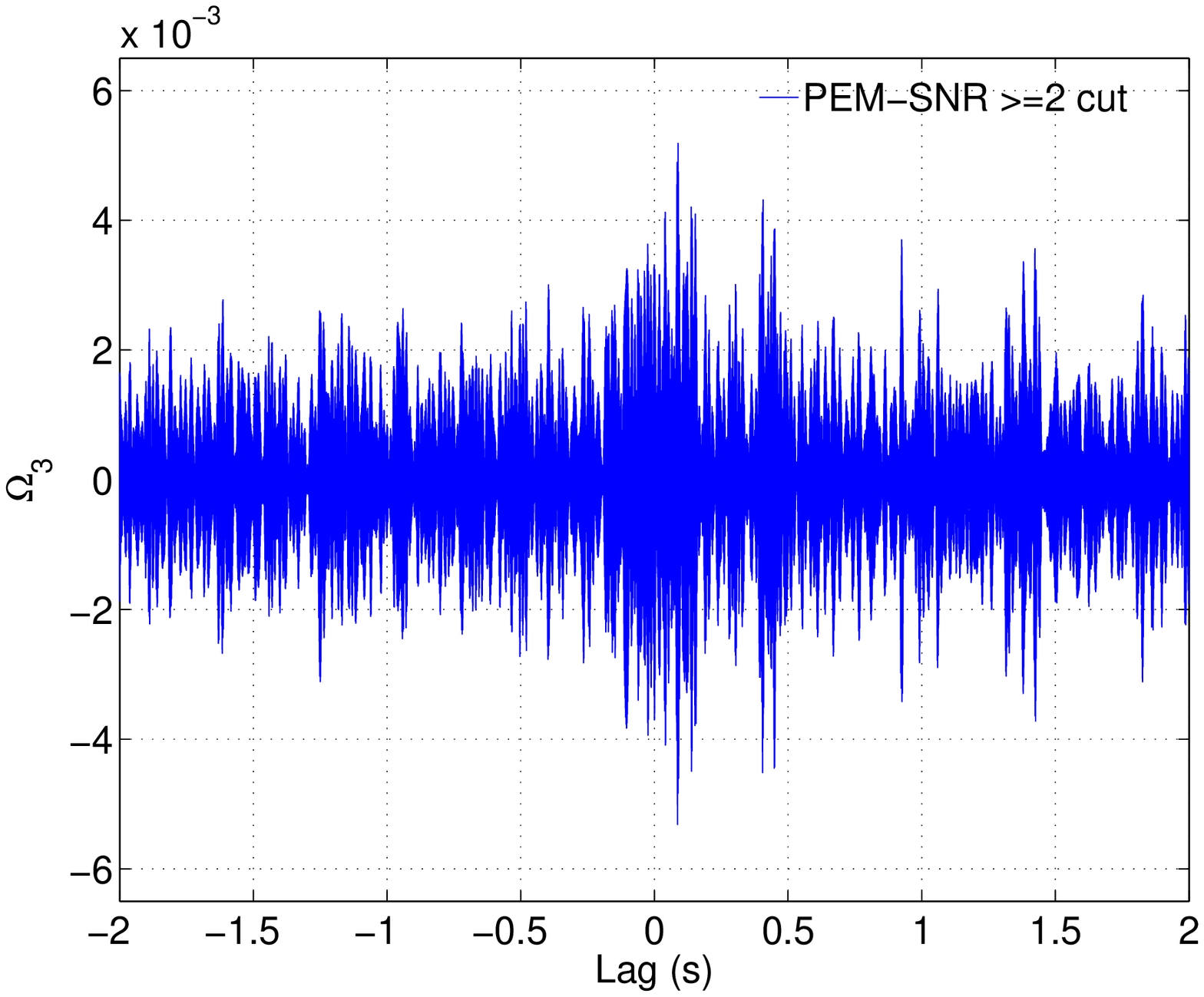}
\\
\includegraphics[angle=0,width=2.6in]{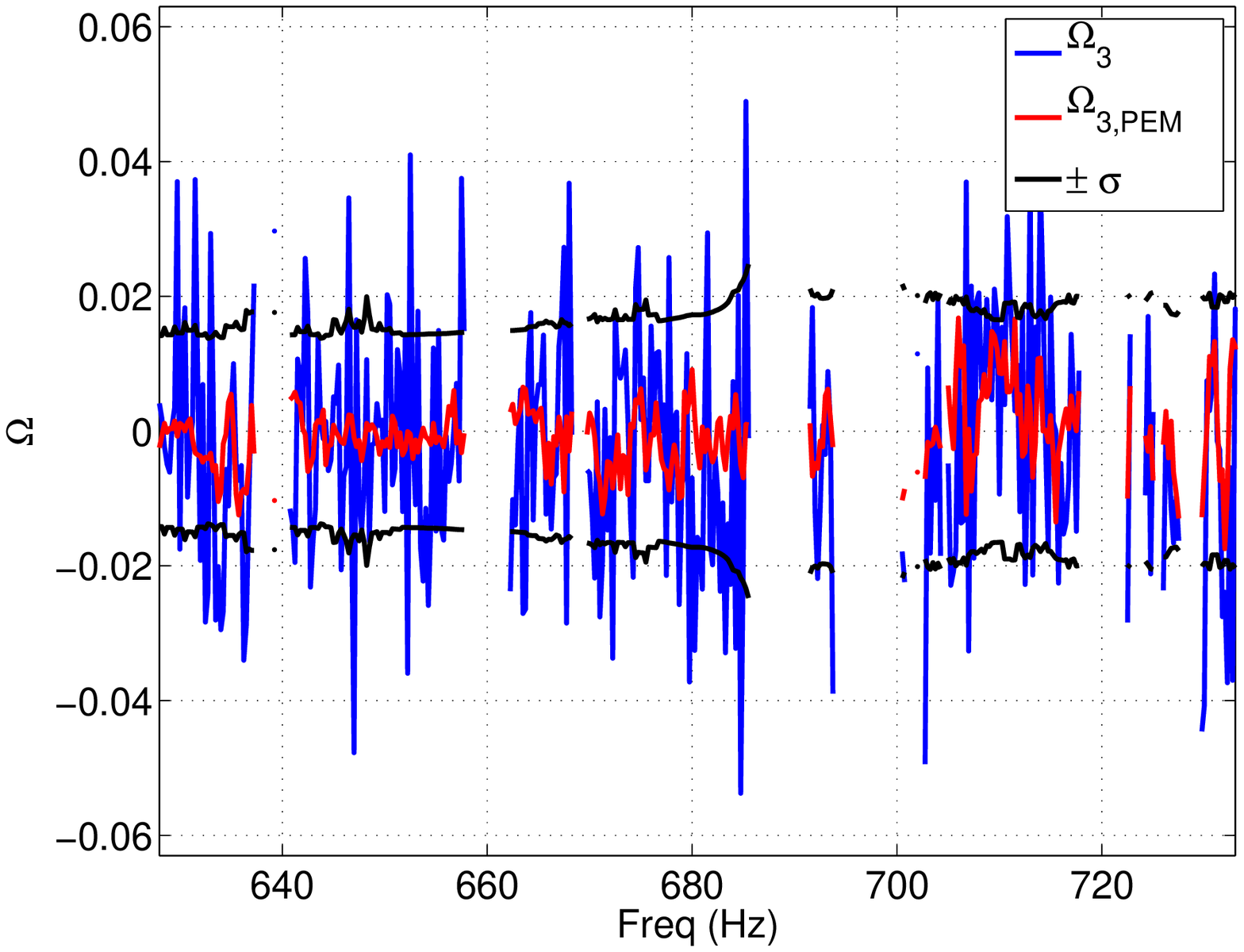}
\includegraphics[angle=0,width=2.6in]{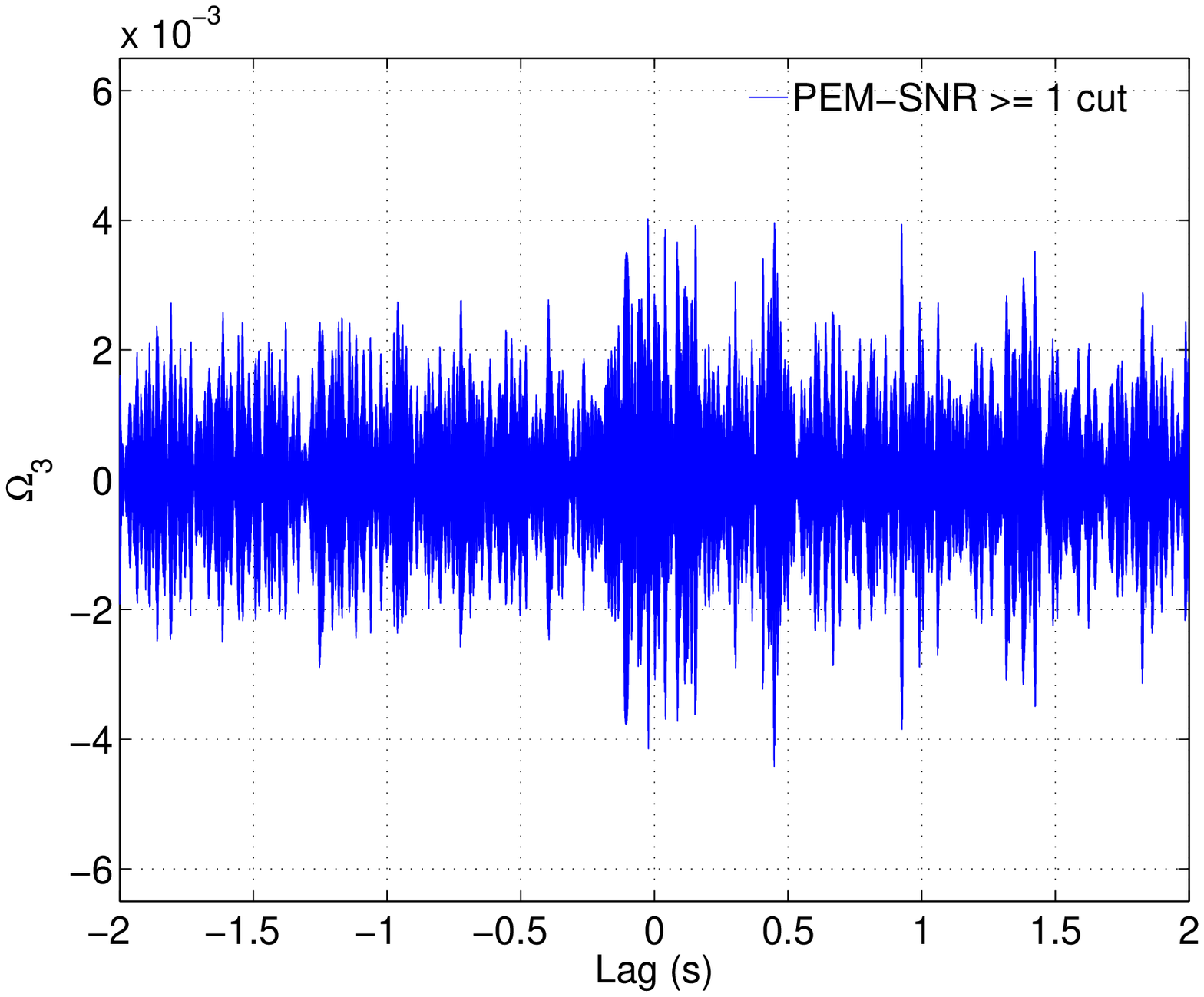}
\end{tabular}
\caption{Plots of $\hat\Omega_3(f)$ and $\hat\Omega_{3,{\rm PEM}}(f)$ (left),
and the inverse Fourier transform of $\hat\Omega_3(f)$ (right), for the
(noisiest) 628--733 Hz sub-band after various stages of noise
removal were applied to the data.
The four rows correspond to the four different stages of cleaning
defined in Table~\ref{t:stages}.
(The top right plot has y-axis limits $13 \times$ greater than the other three.)}
\label{f:HF628}
\end{figure*}

We performed the high-frequency analysis with spectral index $\alpha=3$, and
reference frequency $f_{\rm ref}=900$~Hz. Tables~\ref{t:table1} and
\ref{t:table2} summarize the results after applying several stages of
noise removal as defined in Table~\ref{t:stages}. Table~\ref{t:table1}
applies to the full analysis band, 460--1000~Hz; Table~\ref{t:table2} gives
the results for 5 separate sub-bands. The values of the estimator,
$\hat\Omega_3$, the PEM-coherence contribution to the estimator,
$\hat\Omega_{3,{\rm PEM}}$, and the statistical uncertainty,
$\sigma_{\hat\Omega_3}$, are given for each band and each stage of noise
removal. Also given is the ratio of the standard deviation of the values 
of the inverse Fourier transform of $\hat\Omega_3(f)$ to the statistical 
uncertainty $\sigma_{\hat\Omega_3}$, which is a measure of excess 
residual correlated noise. In the absence of correlated noise, we expect 
the distribution of data points
in the inverse Fourier transform of $\hat\Omega_3(f)$ to follow a
Gaussian distribution with mean 0 and std $\sigma_{\hat\Omega_3}$. Hence
a ratio $\gg 1$ is a sign of excess correlated noise, which shows
up as visible structure in the plot of the inverse
Fourier transform of $\hat\Omega_3(f)$ (for example, see the right hand
plots in Fig.~\ref{f:HF628}). We see that this ratio decreases for the
full 460--1000~Hz band and for each sub-band with every stage of data
cleanup indicating the effectiveness of PEM-coherence SNR cut. We also note
that the values listed in Tables~\ref{t:table1}, \ref{t:table2}
and \ref{t:table3} are the zero lag values of $\hat\Omega_\alpha$ in the
corresponding inverse Fourier transform plots.

\begin{table}
\centering
\begin{tabular}{cccccc}
Stage & $\hat\Omega_3$ & $\hat\Omega_{3,\rm PEM}$ & $\sigma_{\hat\Omega_3}$ &
std/$\sigma_{\Omega_3}$ \\
 & $(\times 10^{-4})$ & $(\times 10^{-4})$ & $(\times 10^{-4})$ & \\
\hline\hline
I & $77.5$ & $-3.05^{\dagger}$ & $2.82$ & 20.5 \\
II & $-2.17$ & $-3.62$ & $3.24$ & 1.18 \\
III & $-4.11$ & $-4.30$ & $3.59$ & 1.04 \\
IV & $-1.29$ & $-2.38$ & $3.64$ & 1.01 \\
\hline\hline
\end{tabular}
\caption{Results for the H1-H2 high-frequency analysis (460--1000~Hz) after
various stages of noise removal were applied to the data. The estimates
$\hat\Omega_3$, PEM-coherence contribution, $\hat\Omega_{3,{\rm PEM}}$ and
$\sigma_{\hat\Omega_3}$ are calculated assuming $H_0=68\ {\rm km/s/Mpc}$.
$\sigma_{\hat\Omega_3}$ is the
statistical uncertainty in $\hat\Omega_3$. The last column gives the ratio of
the standard deviation of the values of the inverse Fourier transform of
$\hat\Omega_3(f)$ to the statistical uncertainty
$\sigma_{\hat\Omega_3}$. As described in Sec.~\ref{sec:highfrequency}, a ratio
much $\gg 1$ is a sign of excess cross-correlated noise.
$^{\dagger}$The PEM-coherence estimate on stage I also excludes frequencies
(including 60 Hz harmonics) and time segments similar to stages II-IV.}
\label{t:table1}
\end{table}

\begin{table}
\centering
\begin{tabular}{cccccc}
Band & Stage & $\hat\Omega_3$ & $\hat\Omega_{3,\rm PEM}$ & $\sigma_{\hat\Omega_3}$ &
std/$\sigma_{\Omega_3}$  \\
(Hz) & & $(\times 10^{-4})$ & $(\times 10^{-4})$ & $(\times 10^{-4})$ & \\
\hline\hline
460--537 & I & $-7.28$ & $-0.22$  & $4.48$ & 5.40 \\
 & II & $-2.17$ & $-0.24$ & $5.08$  & 1.01 \\
 & III & $-0.60$ & $-1.23$ & $5.68$ & 0.98 \\
 & IV & $-0.34$ & $-1.23$ & $5.69$ & 0.97 \\
\hline
537--628 & I & $163$ & $-2.28$ & $5.46$ & 24.0 \\
 & II & $14.7$ & $-2.46$ & $6.32$ & 1.08 \\
 & III & $8.83$ & $-2.00$ & $6.96$ & 1.02 \\
 & IV & $8.56$ & $-1.98$ & $7.03$ & 1.02 \\
\hline
628--733 & I & $512$ & $-16.7$ & $7.33$& 35.9 \\
 & II & $-33.2$ & $-20.5$ & $8.52$ & 1.37 \\
 & III & $-37.0$ & $-16.3$ & $9.20$ & 1.21 \\
 & IV  & $-26.5$ & $-5.88$ & $9.66$ & 1.12 \\
\hline
733--856 & I & $-397$ & $-1.77$ & $8.32$ & 23.0 \\
 & II & $-4.44$ & $-2.24$ & $9.49$ & 1.67 \\
 & III & $-5.29$ & $-6.40$ & $11.0$ & 1.04 \\
 & IV & $2.76$ & $-3.91$ & $11.3$ & 0.98 \\
\hline
856--1000 & I & $89.2$ & $4.63$ & $10.6$ & 3.37 \\
 & II & $2.44$ & $4.63$ & $12.0$ & 1.02 \\
 & III & $0.004$ & $-1.47$ & $13.2$ & 1.01 \\
 & IV & $0.21$ & $-1.41$ & $13.2$ & 1.01 \\
\hline\hline
\end{tabular}
\caption{Same as Table~\ref{t:table1},
but for 5 separate sub-bands of 460--1000~Hz.}
\label{t:table2}
\end{table}

\begin{figure*}
\begin{tabular}{cc}
\includegraphics[angle=0,width=3in]{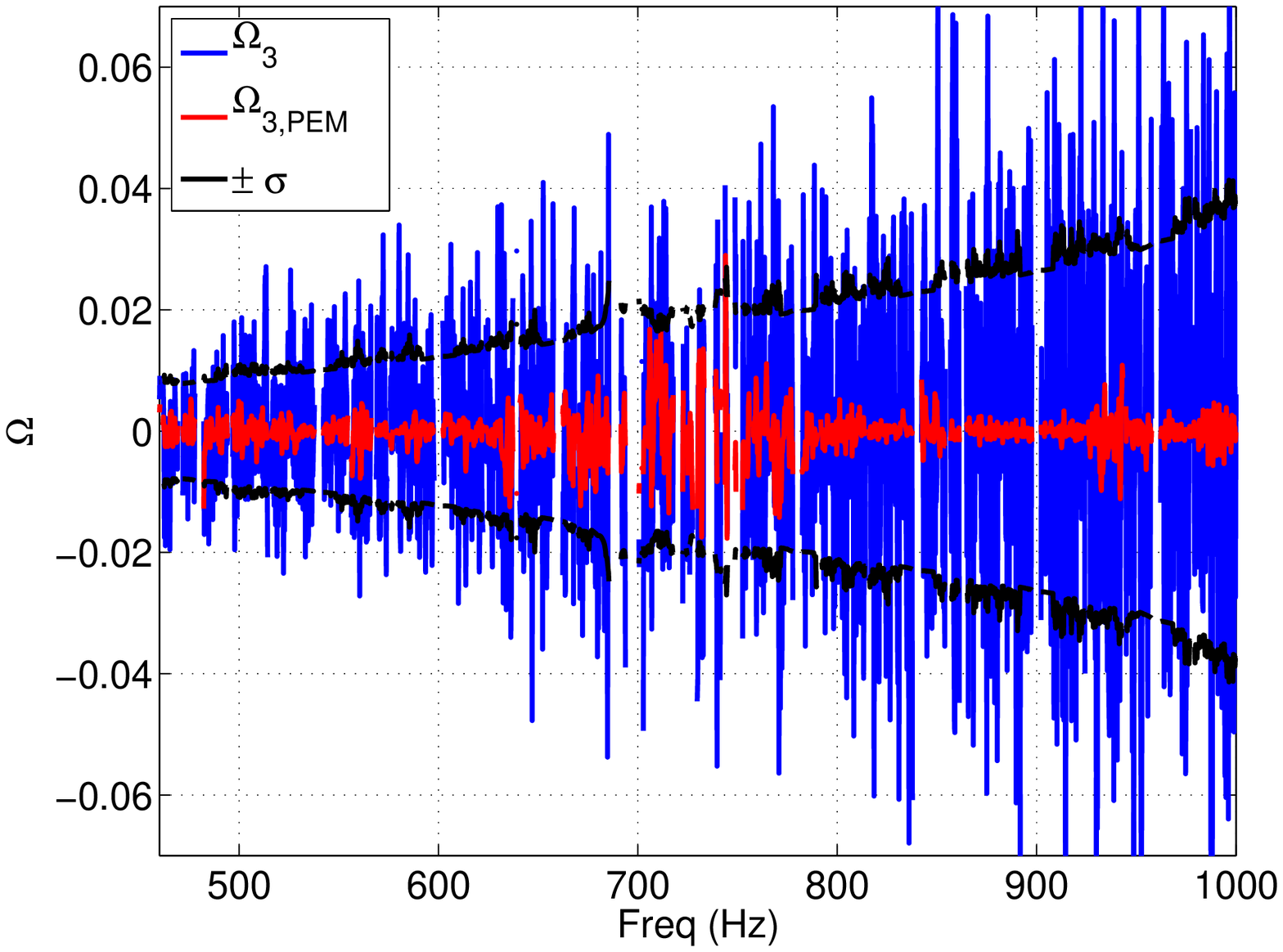}
\includegraphics[angle=0,width=3in]{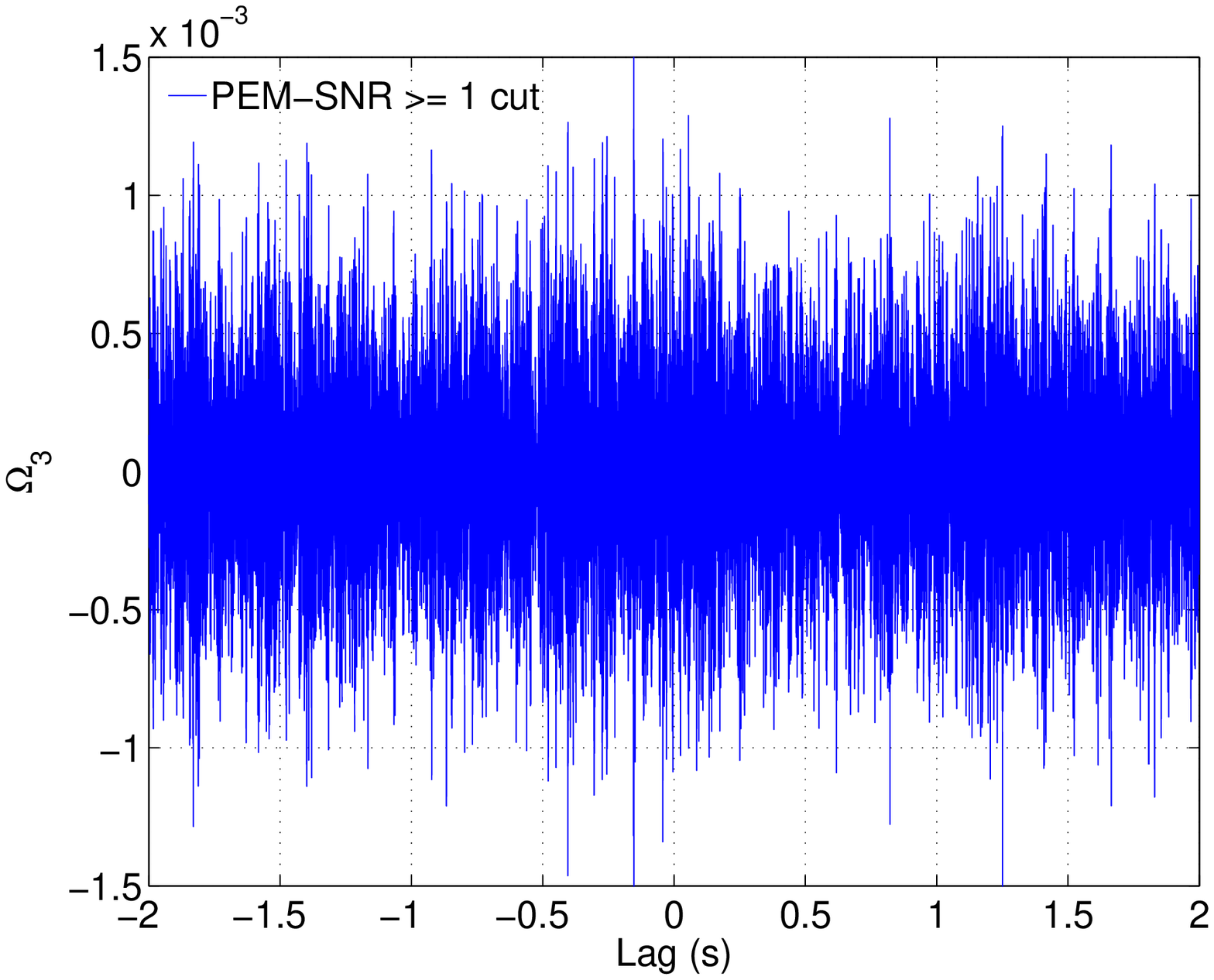}
\end{tabular}
\caption{Plots of $\hat\Omega_3(f)$ and $\hat\Omega_{3,{\rm PEM}}(f)$ (left),
and the inverse Fourier transform of complex $\hat\Omega_3(f)$ (right) for the
full band (460--1000~Hz) after the final stage of noise removal cuts.}
\label{f:HFfull}
\end{figure*}

\begin{figure*}
\begin{tabular}{cc}
\includegraphics[angle=0,width=3in]{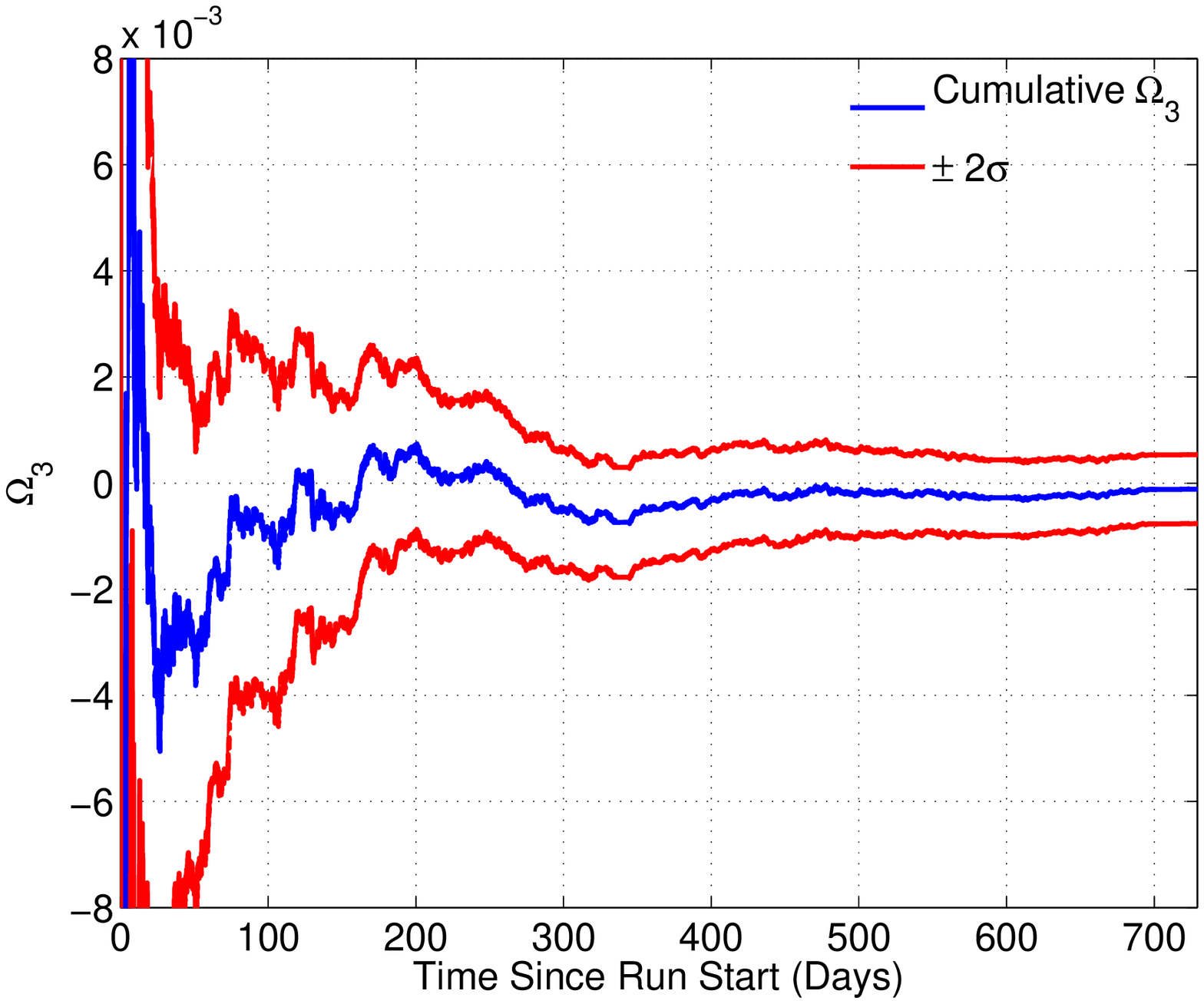}
\includegraphics[angle=0,width=3in]{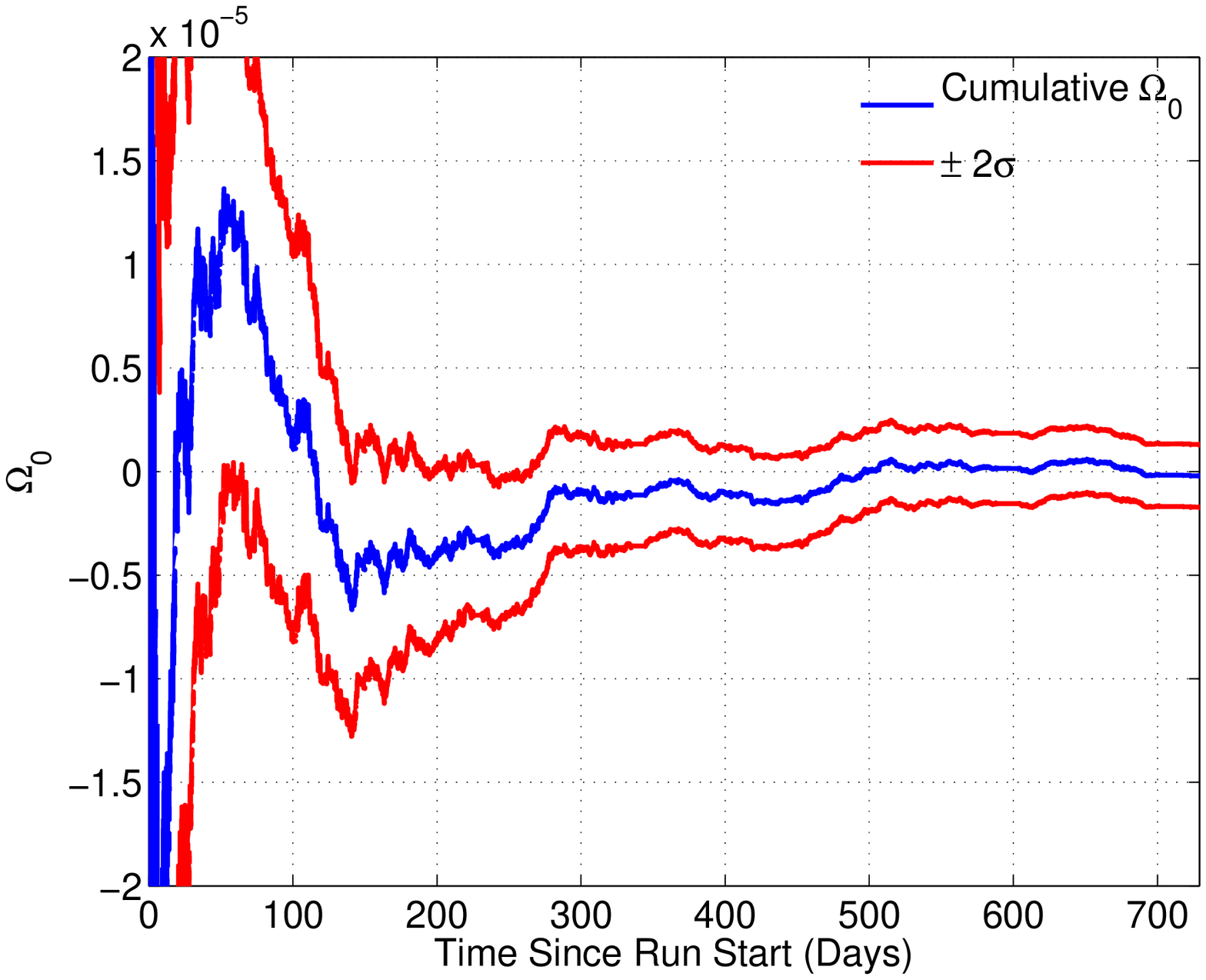}
\end{tabular}
\caption{Running point estimates $\hat\Omega_3$ and $\hat\Omega_0$ for the
high-frequency (460--1000~Hz) and low-frequency (80--160~Hz) analyses,
respectively (left and right panels). The final stage of noise removal cuts
have been applied for both analyses.}
\label{f:rpe}
\end{figure*}

Figure~\ref{f:HF628} is devoted entirely to the noisiest sub-band, 628--733~Hz.
The left column of plots shows $\hat\Omega_3(f)$ and
$\hat\Omega_{3,{\rm PEM}}(f)$, with black lines denoting the statistical error
bar $\pm\sigma_{\hat\Omega_3}(f)$. Here we can clearly see the effectiveness of
noise removal through the four stages discussed above. Note the lack of
structure near zero-lag in the final inverse Fourier transform of the
estimator $\hat\Omega_3(f)$ which is consistent with no correlated noise.
Figure~\ref{f:HFfull} is a similar plot for the full 460--1000~Hz band,
showing the results after the final stage of cuts. Again note the lack of
significant structure near zero-lag in the inverse Fourier transform 
of $\hat\Omega_3(f)$.
Figure~\ref{f:rpe} (left panel) shows how the final estimate, $\hat\Omega_3$,
summed over the whole band, evolves over the course of the run after the final
stage of cuts. The smoothness of that plot (absence of any sharp rise or
fall after the accumulation of sufficient data i.e., one month)
indicates that no particular time period dominates our final result.

\subsection{Low frequency results}
\label{sec:lowfrequency}

\begin{figure*}
\begin{tabular}{cc}
\includegraphics[angle=0,width=2.6in,clip]{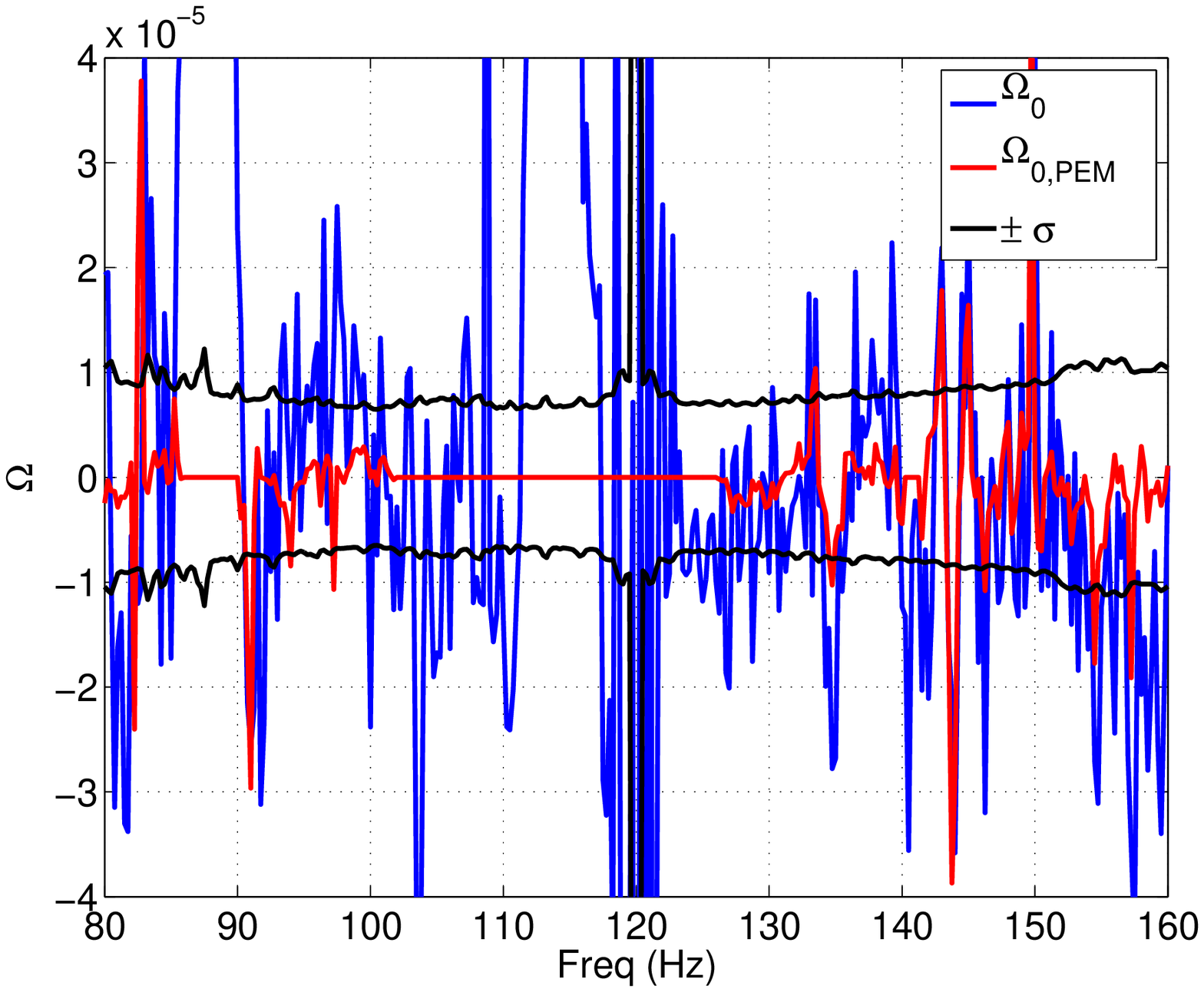}
\includegraphics[angle=0,width=2.6in,clip]{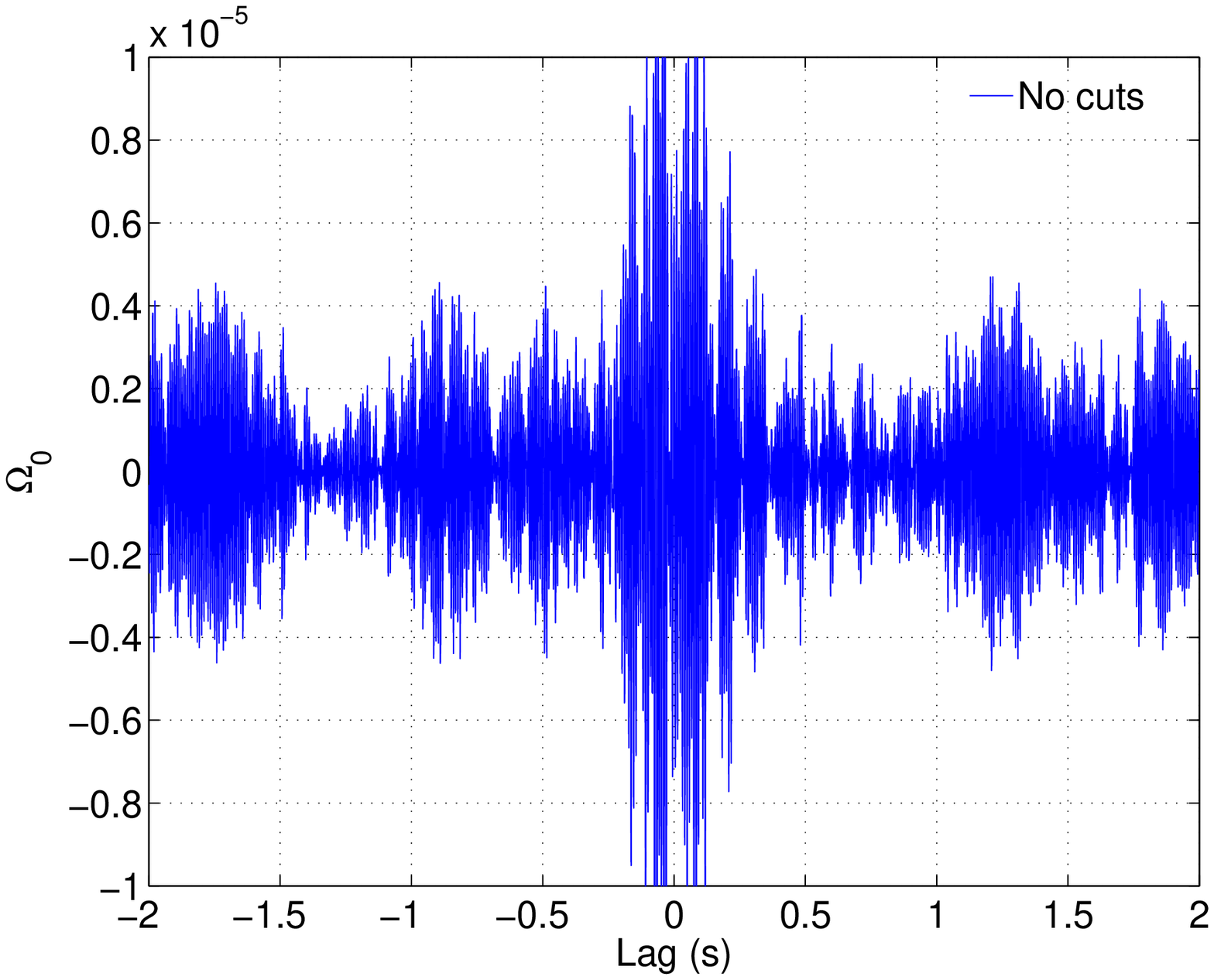}
\\
\includegraphics[angle=0,width=2.6in]{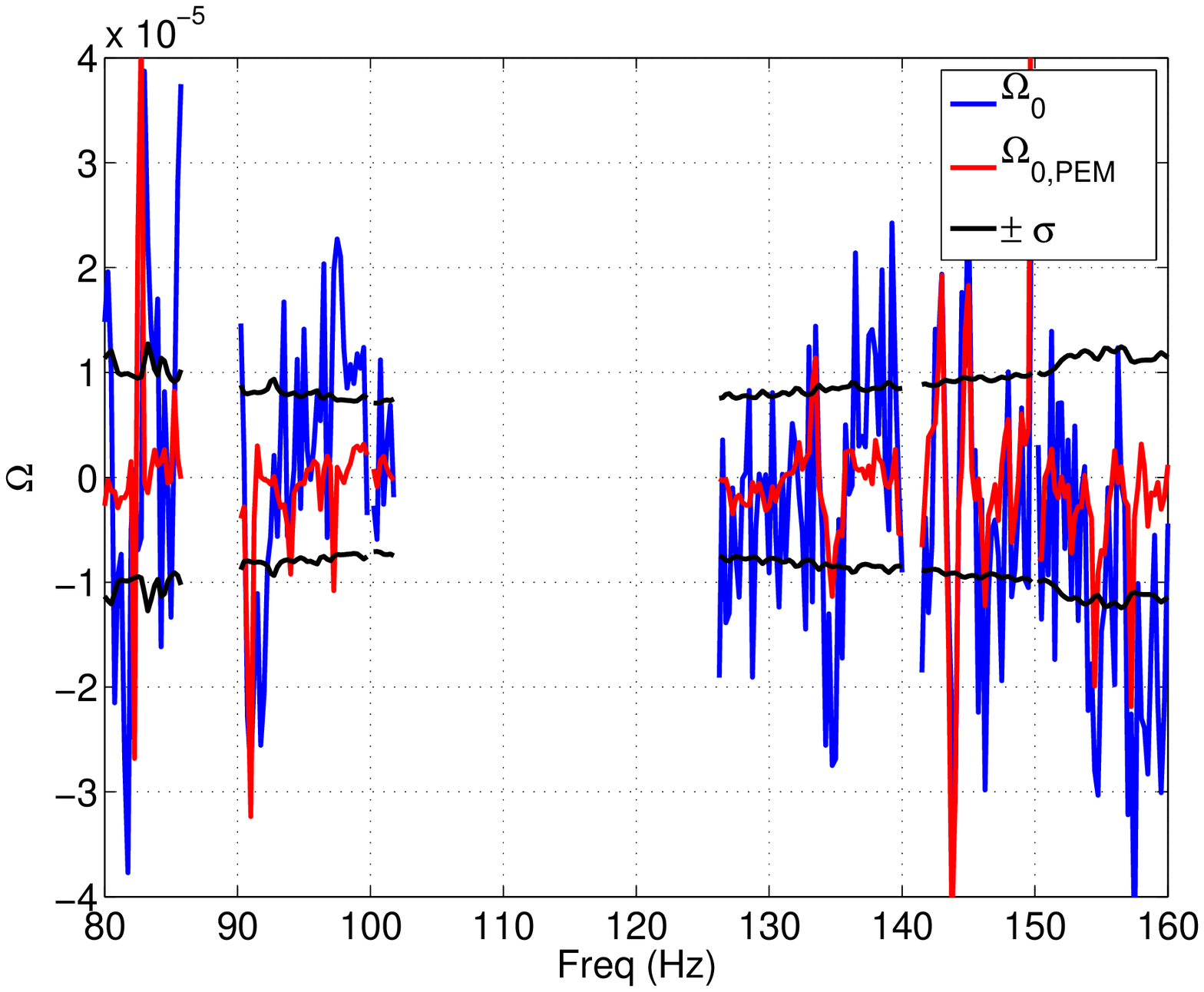}
\includegraphics[angle=0,width=2.6in]{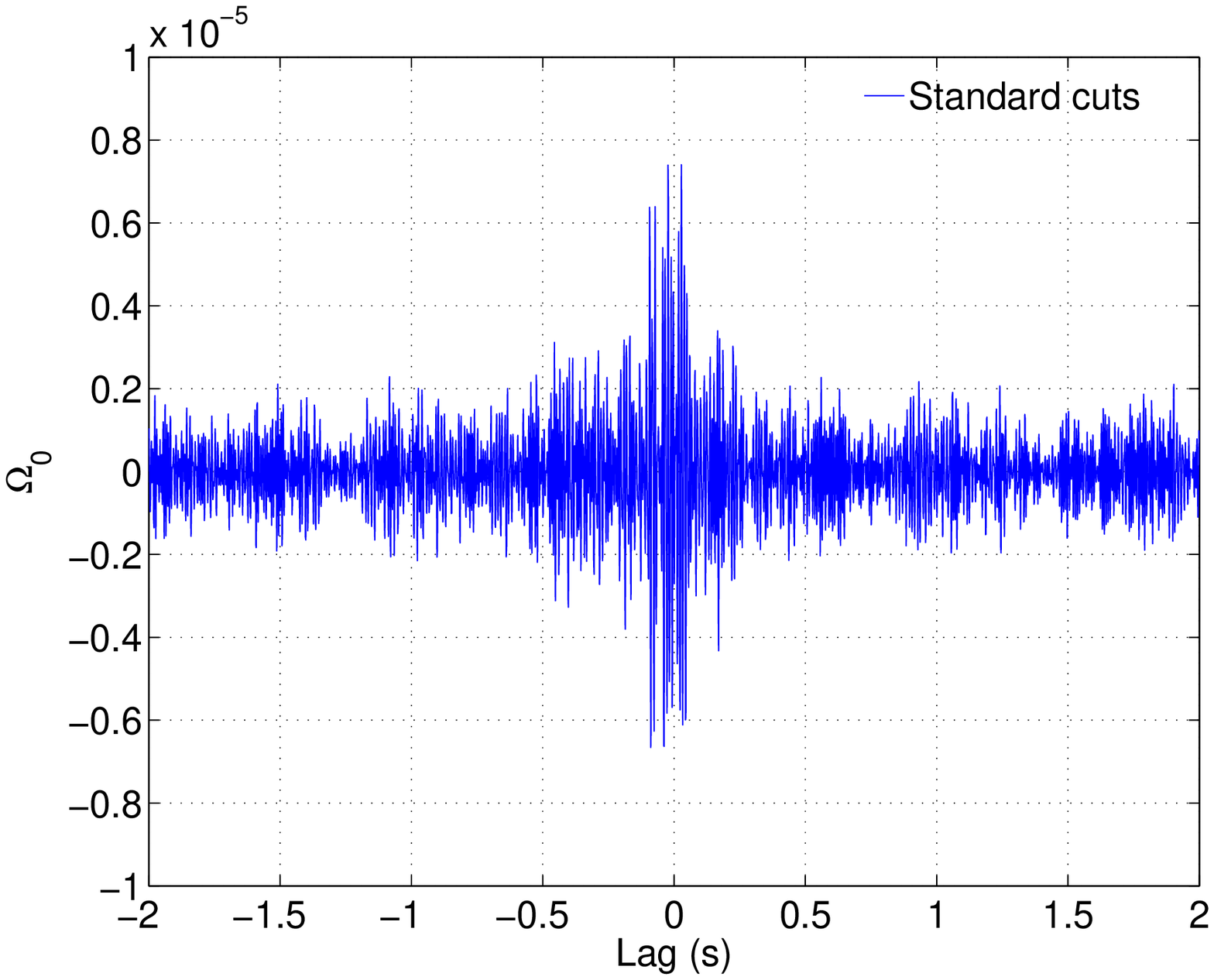}
\\
\includegraphics[angle=0,width=2.6in]{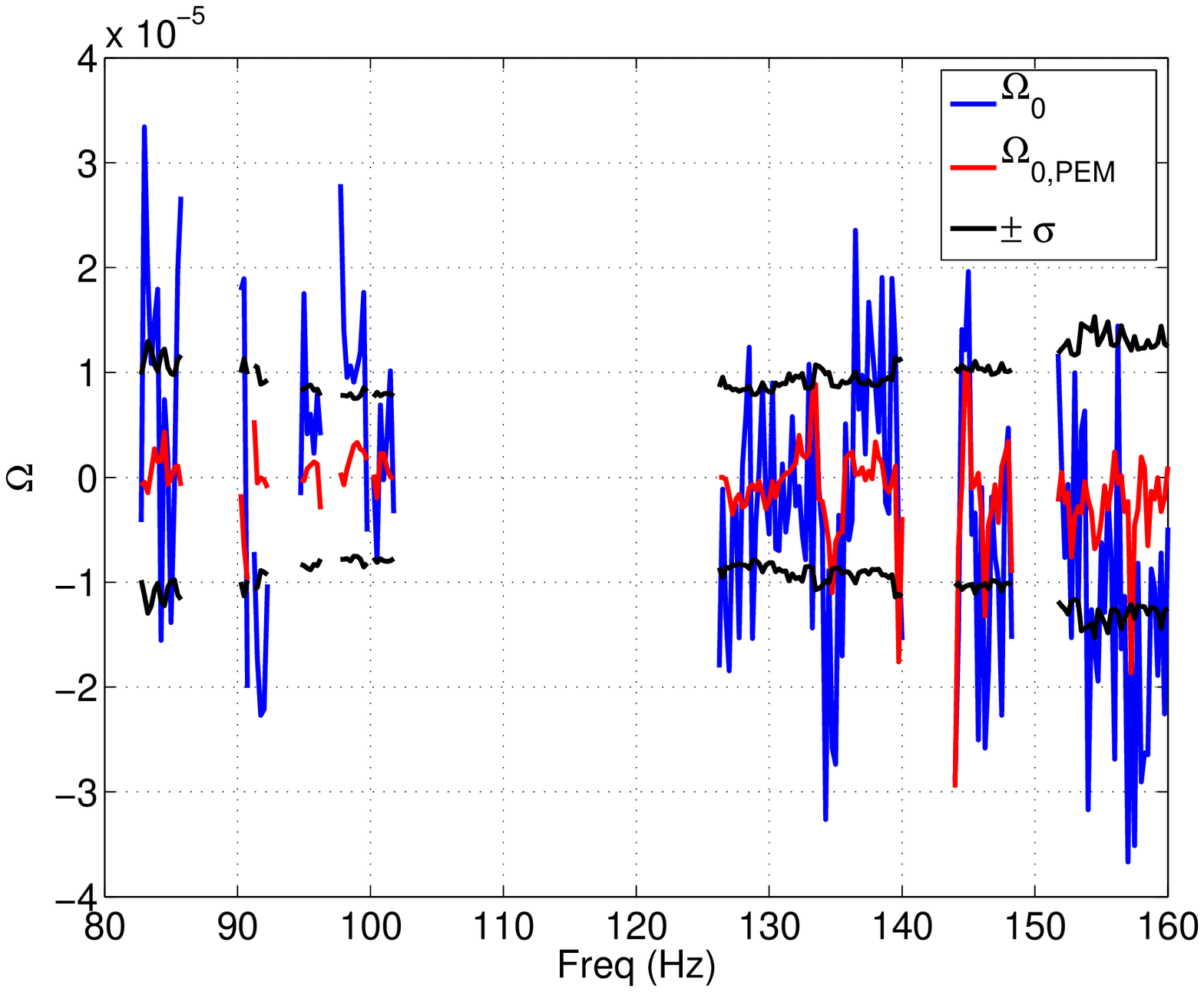}
\includegraphics[angle=0,width=2.6in]{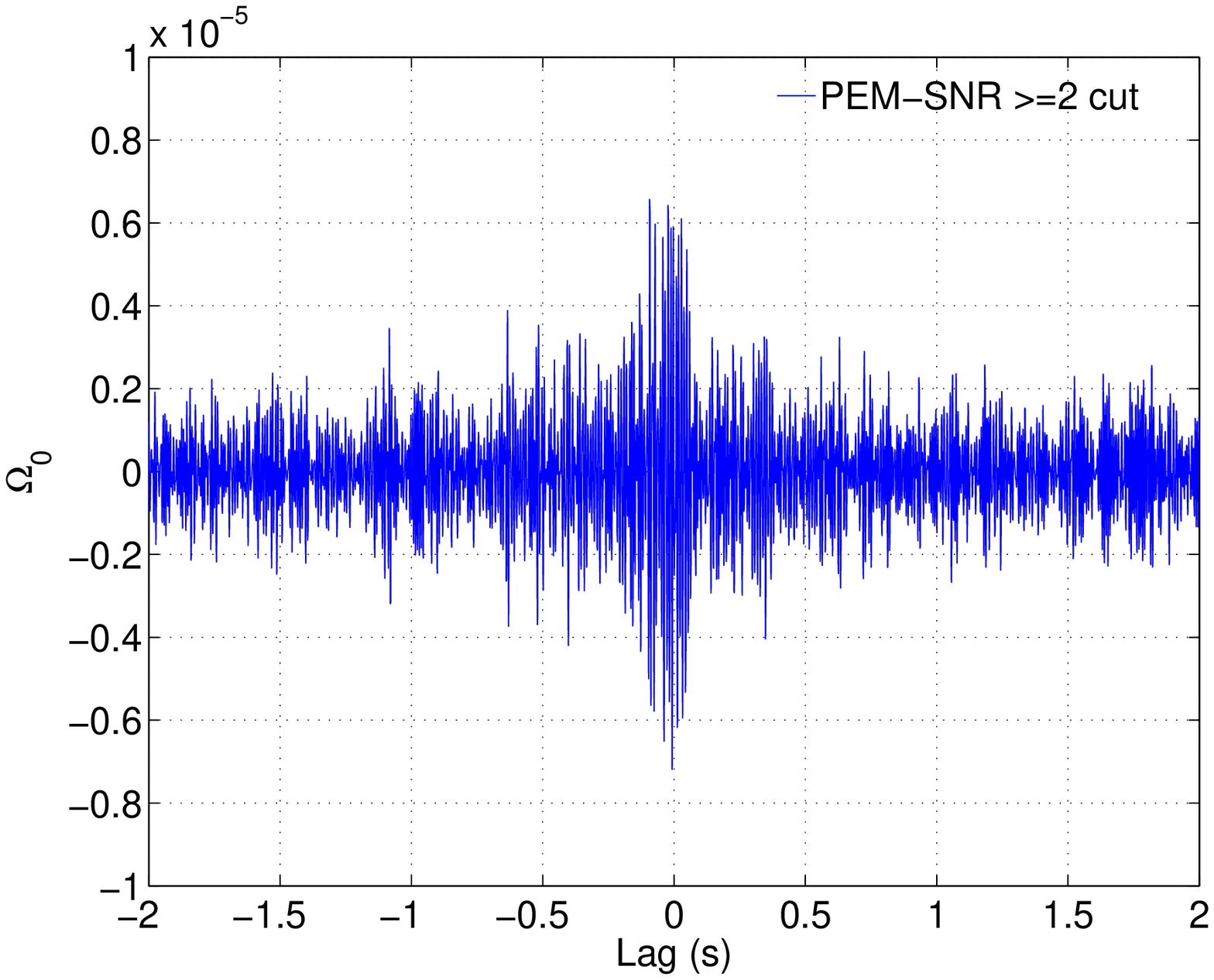}
\\
\includegraphics[angle=0,width=2.6in]{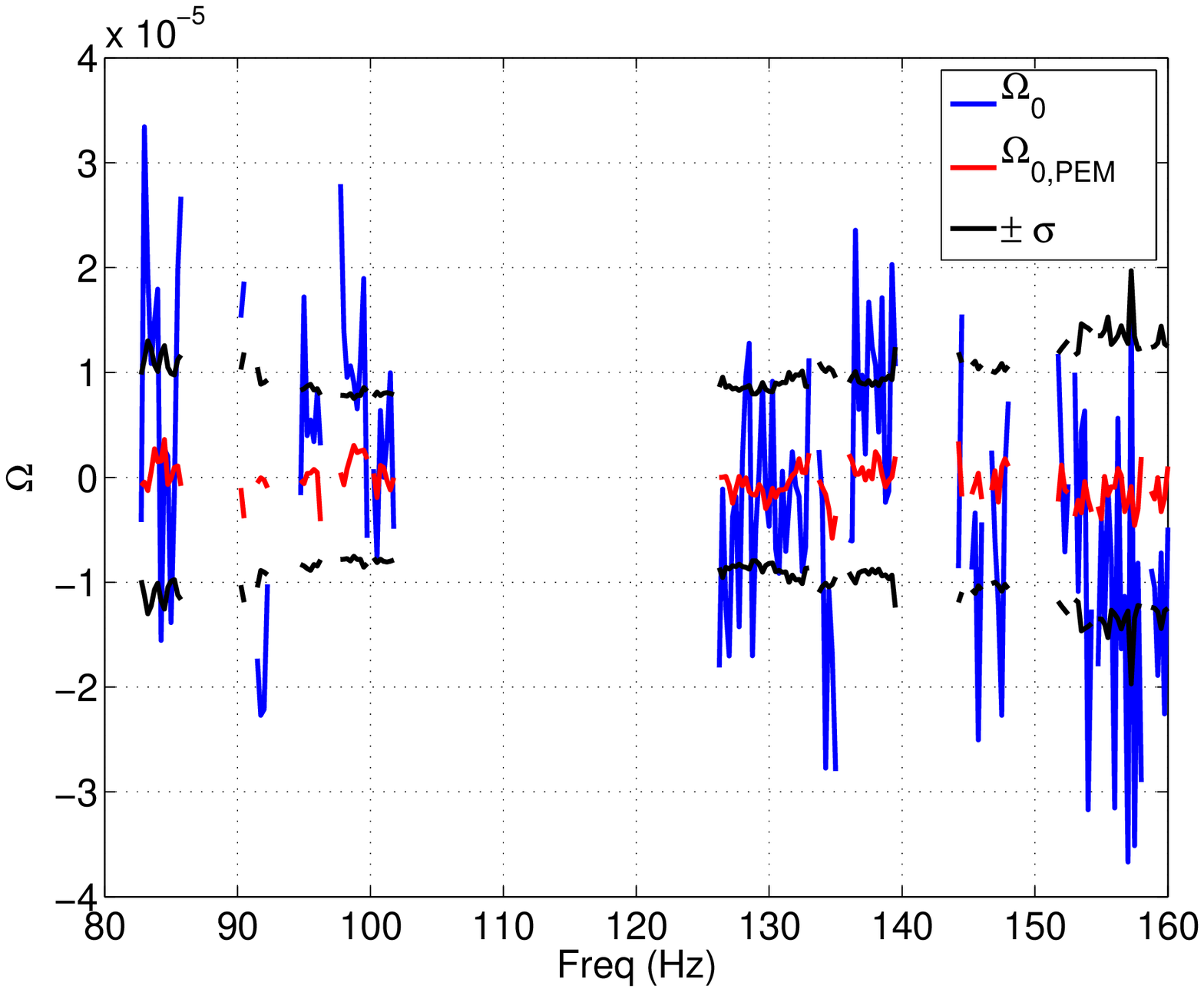}
\includegraphics[angle=0,width=2.6in]{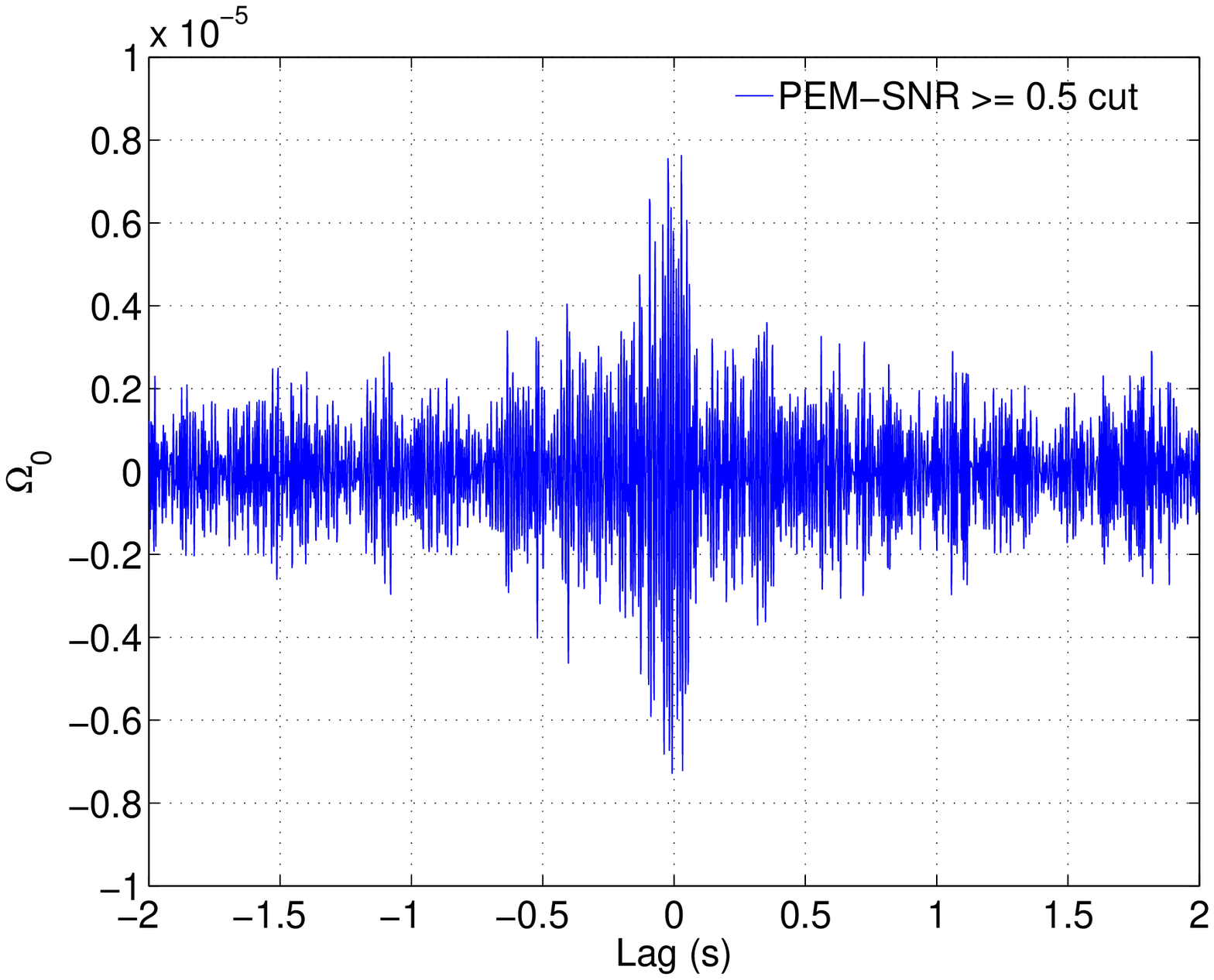}
\end{tabular}
\caption{Plots of $\hat\Omega_0(f)$ and $\hat\Omega_{0,{\rm PEM}}(f)$ (left),
and the inverse Fourier transform of $\hat\Omega_0(f)$ (right) for the
80--160~Hz band after various stages of noise removal were
applied to the data. The four rows correspond to the four different stages of
cleaning defined in Table~\ref{t:stages}.}
\label{f:LF80t160}
\end{figure*}

We now repeat the analysis of the previous subsection but for the
low-frequency band, 80--160~Hz with spectral index $\alpha=0$ and
$f_{\rm ref}=100$~Hz. Table~\ref{t:table3} summarizes the results for 
the low-frequency analysis after applying several stages of noise removal 
as defined in Table~\ref{t:stages}.
\begin{table}
\centering
\begin{tabular}{ccccc}
Stage & $\hat\Omega_0$ & $\hat\Omega_{0,\rm PEM}$ & $\sigma_{\hat\Omega_0}$ &
std/$\sigma_{\Omega_0}$ \\
 & $(\times 10^{-6})$ & $(\times 10^{-6})$ & $(\times 10^{-6})$ & \\
\hline\hline
I & $6.17$ & $-0.39^{\dagger}$ & $0.44$ & $5.90$ \\
II & $-1.71$ & $-0.78$ & $0.63$ & $1.80$ \\
III & $-1.57$ & $-0.84$ & $0.79$ & $1.64$ \\
IV & $-0.26$ & $-0.29$ & $0.85$ & $1.63$ \\
\hline\hline
\end{tabular}
\caption{Similar to Table~\ref{t:table1} but for the low-frequency analysis
(80--160~Hz) and for spectral index $\alpha=0$. The different rows give the
results after various stages of noise removal were applied to the
data. $^{\dagger}$The PEM-coherence estimate on stage I also excludes 
frequencies (including 60 Hz harmonics) and time segments similar to 
stages II-IV.}
\label{t:table3}
\end{table}
Figure~\ref{f:LF80t160} shows the results obtained by applying
the noise removal cuts in four stages. The left
column of plots contain the estimators, $\hat\Omega_0(f)$ and
$\hat\Omega_{0,{\rm PEM}}(f)$, with lines denoting the statistical error bar
$\pm\sigma_{\hat\Omega_0}(f)$.

In contrast to the high-frequency analysis
(compare Figs.~\ref{f:HFfull} and \ref{f:LF80t160}) there is still much
structure in the inverse Fourier transform of $\hat\Omega_0(f)$ around
zero-lag even after the final stage of noise removal cuts were applied.
In addition, the PEM-coherence contribution to the estimator,
$\hat\Omega_{0,{\rm PEM}}(f)$, displays much of the structure observed
in $\hat\Omega_0(f)$. Both of these observations suggest contamination from
residual correlated instrumental or environmental noise
that was not excluded by the noise removal methods. Figure~\ref{f:rpe}
(right panel) shows how the final estimate, $\hat\Omega_0$, evolves over
the course of the run after the final stage of cuts. We note here that
even though $\hat\Omega_0$ (last entry in Table~\ref{t:table3}) is
consistent with zero (within $2\sigma$), its estimate at other non-zero lags
vary strongly as shown in Fig.~\ref{f:LF80t160} (lower right). This indicates
the presence of residual correlated noise after all the time-shift and
PEM-coherence noise removal cuts are applied.

\subsection{Hardware and software injections}
\label{sec:injections}

\begin{figure*}
\includegraphics[width=3in]{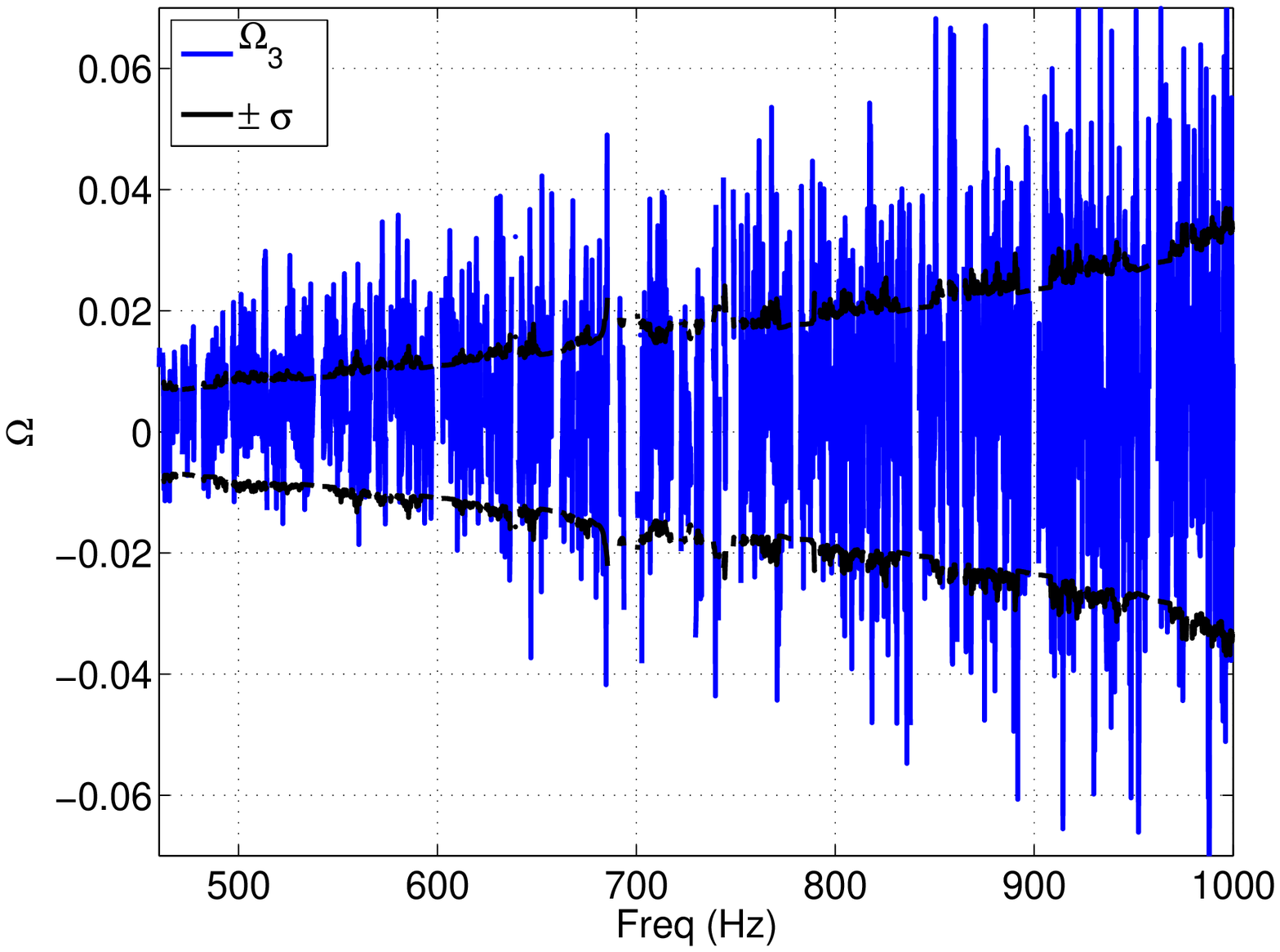}
\includegraphics[width=3in]{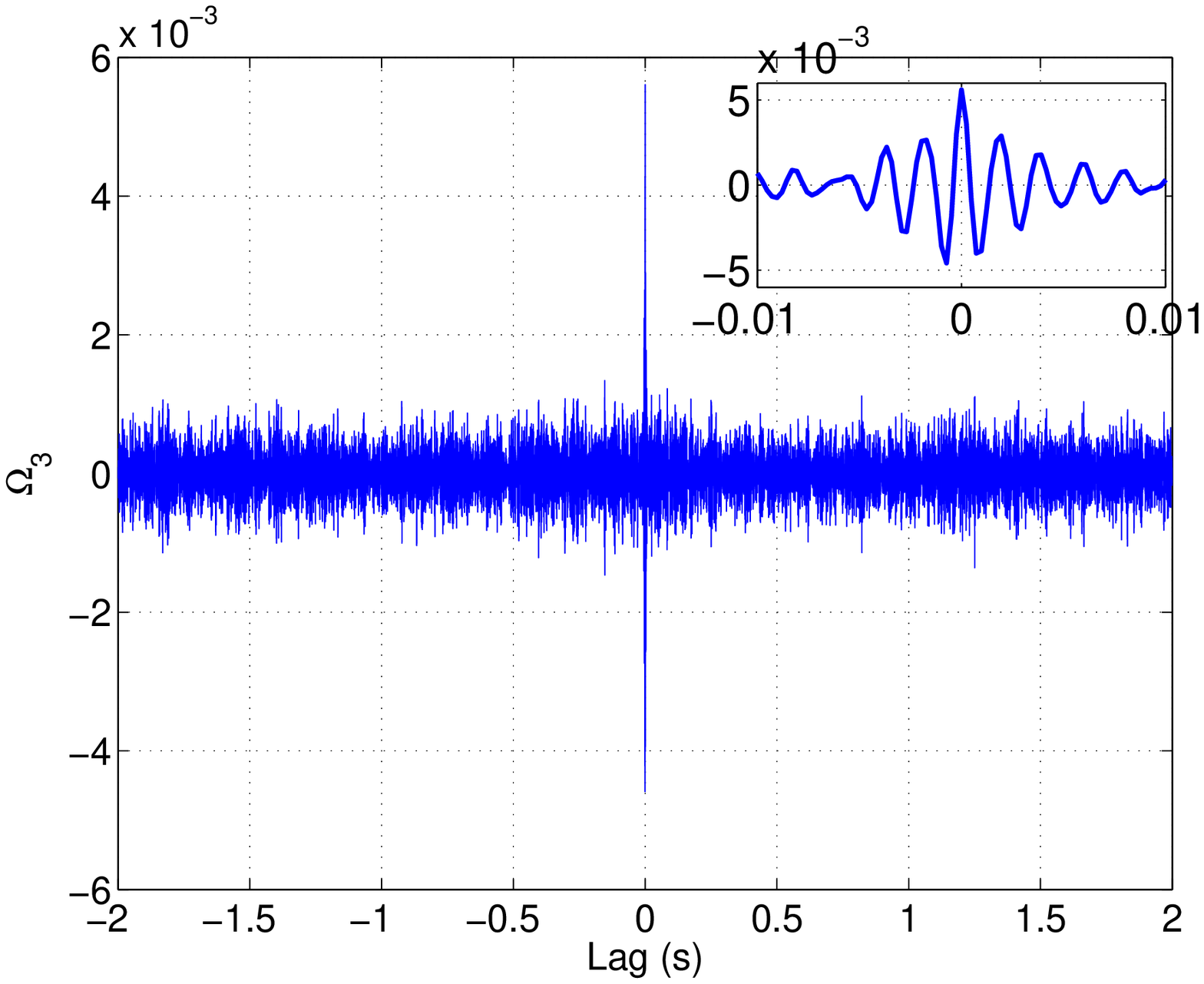}
\caption{Left panel: Recovered spectrum for a software injection with an
amplitude $\Omega_3 = 5.6 \times 10^{-3}$ (SNR $\sim$ 17). Right panel: The
inverse Fourier transform of the recovered $\hat\Omega_3(f)$ and a $\pm 10$ms
zoom-in around zero lag.}
\label{softinj}
\end{figure*}

We validated our analysis procedure by injecting simulated stochastic GW
signals into the strain data of the two detectors. Both {\em hardware} and
{\em software} injections were performed.
Hardware injections are performed by physically moving the interferometer
mirrors coherently between interferometers. In this case the artificial
signals were limited to short durations and relatively large amplitudes.
The data from these hardware injection times were excluded from the analyses
described above, as noted in Sec.~\ref{sec:steps}, Step 1.
Software injections are conducted by adding a simulated GW signal to the
interferometer data, in which case they could be long in duration and
relatively weak in amplitude. During S5 there was one stochastic signal
hardware injection when both H1 and H2 were operating in coincidence.
A stochastic background signal with spectral index $\alpha=0$ and amplitude
$\Omega_0 = 6.56 \times 10^{-3}$ was injected for approximately 3~hours.
In performing the analysis, frequency bins were excluded based on the
standard H1-H2 coherence calculations. No additional frequency bins were
removed using ${\rm SNR}_{\rm PEM}$.
The recovered signal was $\Omega_0 = (7.39 \pm 1.1) \times 10^{-3}$, which is
consistent with the injected amplitude. Due to the spectral
index used for the injection ($\alpha = 0$), the recovery analysis was
performed using only the low frequency band. We also performed a software 
injection in the high frequency band with an amplitude 
$\Omega_3 = 5.6 \times 10^{-3}$, and we recovered it successfully. 
Figure~\ref{softinj} shows the spectrum of the recovered
$\hat\Omega_3(f)$ and its inverse Fourier transform.

\section{Assessing the residual correlated noise}
\label{sec:upperlimits}

After applying the full noise removal procedure, the high-frequency band
appears clean whereas the low-frequency band exhibits evidence of residual
correlated noise. In order to interpret the implications of these two very 
different results, we introduce a general procedure for determining whether a
stochastic measurement is sufficiently well-understood to yield an
astrophysical interpretation. While our immediate concern is to provide a 
framework for interpreting the two
results presented here, we aim to give a comprehensive procedure that can be
applied generally, to both co-located and non-co-located detectors.
In this spirit, this section is organized as follows: first, we present a
general framework for interpreting stochastic measurements;
then we discuss how it can be applied to (familiar) results from
non-co-located detectors; and finally we apply the framework to our 
present results.

To determine whether a result can be interpreted as a constraint
on the SGWB, we consider the following three criteria:
\begin{enumerate}
  \item We have accounted for all known noise
sources through either direct subtraction,
vetoing, and/or proper estimation of systematic errors.

  \item Having accounted for known noise sources,
we do not observe evidence of residual noise that
is inconsistent with our signal and noise models.

  \item To the best of our knowledge, there is no
plausible mechanism by which broadband correlated noise
might be lurking beneath the uncorrelated noise at a
level comparable to the GW signal we are trying to measure.
\end{enumerate}
If an analysis result does not meet these criteria, then we conservatively 
place a bound on the sum of the GW signal and the residual correlated noise.
If a result meets all the criteria, then we present astrophysical bounds on 
just the GW signal.

Let us now examine these criteria in the context of previous results using 
the non-co-located LHO and LLO detectors~\cite{S5HLiso}. Criterion~\#1 was 
satisfied by identifying and removing instrumental lines attributable to 
known instrumental artifacts such as power lines and violin resonances.
Criterion~\#2 was satisfied by creating diagnostic plots, e.g., showing 
$\hat\Omega_0$ vs.\ lag (the delay time between the detectors; 
see~Fig.~\ref{f:HF628}), which demonstrated that the measurement was 
consistent with uncorrelated noise (and no GW signal). Criterion~\#3 was 
satisfied by performing order-of-magnitude calculations for plausible 
sources of correlated noise for LHO-LLO including electromagnetic phenomena, 
and finding that they were too small to create broadband correlated noise 
at a level that is important for initial LIGO.

Next, we consider how the criteria might be applied to future measurements 
with non-co-located detectors. During the advanced detector era, correlated 
noise from Schumann resonances may constitute a source of correlated noise 
at low frequencies $\lesssim\unit[200]{Hz}$, even for widely separated 
detector pairs such as LHO-LLO~\cite{schumann,wsubtract}. While it may be 
possible to mitigate this potential correlated noise source through 
commissioning of the detectors to minimize magnetic coupling, or failing 
that, through a noise subtraction scheme, we consider the possibility that 
residual correlated noise is observed. In this scenario, we could still aim 
to satisfy criteria~\#2 and \#3 by using magnetometer measurements to 
construct a correlated noise budget, which could then be used to interpret 
the results.

Finally, we consider how the criteria apply to the measurements presented
in this paper. The high-frequency analysis meets criteria 1 and 2 as we did not
observe residual noise inconsistent with our noise models
(see~Fig.~\ref{f:HFfull}). We did observe residual noise for the low frequency
analysis (see~Fig.~\ref{f:LF80t160}), but it was consistent with a preliminary
noise model, based on measured acoustic coupling and microphone signals
(most of the channels identified by the PEM coherence method were
either microphones or accelerometers placed on optical tables that
were susceptible to acoustic couplings). While
the bands that were acoustically loudest (containing certain electronics
fans) were vetoed, the acoustic coupling in between the vetoed bands was
high enough to produce a residual signal. We did not further develop the
noise model to meet criterion 1 because, with the systematic error from
acoustic coupling, the astrophysical limit would not have improved on
values we have reported previously \cite{S5HLiso,S6-VSR23}. For this reason,
we do not present an astrophysical limit for the low frequency band.

We addressed criterion \#3 in two ways. First, by investigating mechanisms
that might produce un-monitored broad-band correlations between
detectors, such as the study of correlations introduced by the shared data 
acquisition system, the study of correlations introduced by light 
scattering, and PEM coverage studies described in Sec.~\ref{sec:othersources}.

We also identified the sources of most of the features between 80 and 400~Hz.
For many of the spectral peaks, in addition to coherence between the
GW channels, there was also coherence between the individual GW channels and
the accelerometer and microphone signals from the vertex area shared by 
both detectors. The coupling was consistent with the measured coupling of 
acoustic signals to the detectors. Most of these features were traced to 
electronics cooling fans in specific power supply racks in the vertex 
station by comparing coherence spectra to spectra for accelerometers 
mounted temporarily on each of the electronics racks. The features were 
produced at harmonics of the fan rotation frequencies.

The second type of coherence feature was associated with bilinear coupling of 
low frequency ($<15$~Hz) seismic motion and harmonics of 60~Hz, producing 
side-band features around the harmonics that were similar to the features 
in the 0--15~Hz seismic band. Coherence of side-band features was expected 
since the coherence length of low-frequency seismic signals was greater
than the distance separating sensitive parts of the two interferometers at 
the vertex station, and the seismic isolation of the interferometers was 
minimal below 10~Hz.

In conclusion, we found no peaks or features in the coherence spectrum for 
the two GW channels that were inconsistent with linear acoustic coupling 
or bilinear coupling of low frequency seismic noise and 60~Hz harmonics 
at the vertex station. Neither of these mechanisms is capable of producing
broad-band coherence that is not well monitored by the PEM system.
Therefore, for the high frequency analysis, we satisfy the three criteria
for presenting astrophysical bounds on just the GW signal.

\subsection{Upper-limits}
\label{sec:hfUL}

Since there is no evidence of significant residual noise contaminating the 
high-frequency data after applying the full set of cuts, we set a 95\% 
confidence-level Bayesian upper-limit on $\Omega_3$. We use the previous 
high-frequency upper limit $\Omega_3 < 0.35$ (adjusted for 
$H_0 = 0.68$ km/s/Mpc) from the LIGO S5 and Virgo VSR1 
analysis~\cite{S5-VSR1-LIGO-Virgo} as a prior and assume a flat distribution 
for $\Omega_3$ from 0 to 0.35. We also marginalize over the calibration 
uncertainty for the individual detectors (10.2\% and 10.3\% for H1 and 
H2, respectively). In order to include in our calculation the PEM estimate 
of residual contamination, we take 
$\sigma_{\hat\Omega_3}^2 + \hat\Omega_{3,PEM}^2$ as our
total variance. We note here that the estimated $\hat\Omega_{3,PEM}$ is
within the observed $\sigma_{\hat\Omega_3}$ i.e., we observe no evidence of
excess environmental contamination and the above quadrature addition increases
the limit by $\sim 20$\%. The final result is 
$\Omega_3 < 7.7 \times10^{-4}$ for the frequency band 460-1000
Hz, which is an improvement by a factor of $\sim\!180$ over the 
recent S6/VSR2-3 result~\cite{S6-VSR23}. All of the above $\sim\!180$ factor 
improvement comes from the nearly-unity overlap reduction function of the 
co-located Hanford detectors. In fact, all other data being same, if we were to 
consider the H2 detector to not be located at Hanford but instead at the LIGO 
Livingston site yields an upper limit that is worse by a factor of $\sim\!1.7$ than 
the S6/VSR2-3 result. Most of this difference of $\sim\!1.7$ comes from the 
improved sensitivities of S6/VSR2-3 detectors compared to S5 H1-H2 detectors. 
Upper-limits for the five separate 
sub-bands of the high-frequency analysis are given in Table~\ref{t:ULhighfreq}.
\begin{table}[h!]
\centering
\begin{tabular}{c|c}
Band (Hz) & 95\% CL UL $(\times 10 ^{-3})$ \\
\hline
460--1000 & 0.77 \\
\hline
460--537 & 1.11 \\
537--628 & 2.12 \\
628--733 & 1.18 \\
733--856 & 2.53 \\
856--1000 & 2.61 \\
\hline\hline
\end{tabular}
\caption{95\% confidence level upper-limits for the the full band 
(460--1000~Hz) and for five separate sub-bands.}
\label{t:ULhighfreq}
\end{table}

As mentioned in Sec.~\ref{sec:lowfrequency}, the structure in the inverse 
Fourier transform plots of Fig.~\ref{f:LF80t160} suggests contamination from 
residual correlated noise for the low-frequency analysis and hence we do not 
set any upper-limit on $\Omega_0$ using the low-frequency band 80-160~Hz.

\section{Summary and plans for future analyses}
\label{sec:summary}

In this paper, we described an analysis for a SGWB using data taken by the 
two co-located LIGO Hanford detectors, H1 and H2, during LIGO's fifth 
science run. Since these detectors share the same local environment,
it was necessary to account for the presence of correlated instrumental and 
environmental noise. We applied several noise identification and mitigation
techniques to reduce contamination and to estimate the bias due to any 
residual correlated noise. The methods proved to be useful in cleaning the 
high-frequency band, but not enough in the low-frequency band.

\begin{figure}
\includegraphics[width=3.5in]{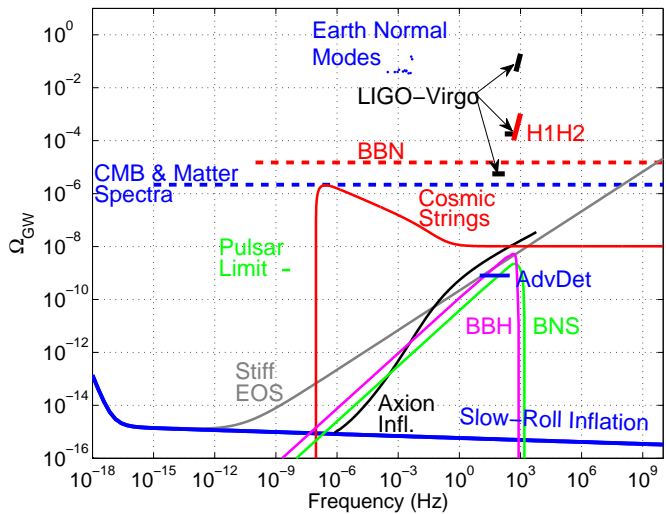}
\caption{(Color online) Upper limits from the current H1-H2 analysis, 
previous SGWB analyses
and the projected advanced LIGO limit, along with various SGWB models.The BBN
limit is an integral limit on $\Omega_{\rm gw}$ i.e.,
$\int \Omega_{\rm gw} (f) d(lnf)$ in the $10^{-10} - 10^{10}$  Hz band
derived from the Big Bang Nucleosynthesis and observations of the abundance
of light nuclei \cite{BBNResult,S5HLiso}. The measurements of CMB and matter
power spectra provide a similar integral bound in the frequency range of
$10^{-15} - 10^{10}$ Hz \cite{CMBresult}. The pulsar limit is a bound on
the $\Omega_{\rm gw} (f)$ at $f = 2.8$ nHZ and is based on the fluctuations
in the pulse arrival times from millisecond pulsars \cite{Pulsar_result}. In
the above figure, slow-roll inflationary model \cite{Inflation}
assumes a tensor-to-scalar ratio of $r = 0.2$, the best fit value from the
BICEP2 analysis \cite{BICEP2}. In the axion based inflationary model, for
certain ranges
of parameters the backreaction during the final stages of inflation is expected to produce strong GWs at high frequencies \cite{axion_inflation}. The stiff
equation of state (EOS) limit corresponds to scenarios in the early universe
(prior to BBN) in which GWs are produced by an unknown `stiff'
energy \cite{stiffEOS}. For the above figure we used the equation of
state parameter $w = 0.6$ in stiff EOS model. The cosmic string model
corresponds to GWs produced by cosmic strings in the early universe
\cite{cosmic_string_result}. The Earth's normal mode limits are based on the
observed fluctuations in the amplitudes of Earth's normal modes using an
array of seismometers \cite{Earthmodes}. The astrophysical SGWBs (BBH and BNS)
are due to the superposition of coalescence GW signals from a large number of
binary black holes (BBH) and binary neutron stars (BNS) \cite{SGWBBinary}.}
\label{f:landscape}
\end{figure}

In the $\unit[80-160]{Hz}$ band, we were unable to sufficiently mitigate the
effects of correlated noise, and hence we did not set any limit on
the GW energy density for $\alpha=0$.
For the $\unit[460-1000]{Hz}$ band, we were able mitigate the effects of
correlated noise, and so we placed a  95\% C.L.~upper limit on the GW energy
density alone in this band of $\Omega_3 < 7.7\times 10^{-4}$.
This limit improves on the previous best limit in the high-frequency band
by a factor of $\sim 180$~\cite{S6-VSR23}. Figure~\ref{f:landscape}
shows upper limits from current/past SGWB analyses,
as well as limits from various SGWB models, and
projected limits using Advanced LIGO. We note here that the indirect limits
from BBN apply to SGWBs present in the early universe at the time of BBN 
(and characterized by an $\alpha=0$ power law; 
see Eq.~\ref{e:Omega_gw_alpha}), but not to SGWBs of
astrophysical origin created more recently (and assumed to be characterized
by an $\alpha=3$ power law).
Thus, the results presented here complement the indirect bound from BBN,
which is only sensitive to cosmological SGWBs from the early universe,
as well as direct $\alpha=0$ measurements using lower-frequency observation bands \cite{S5HLiso}.

There are several ways in which the methods presented in this paper can be 
improved. We list some ideas below:

(i) As mentioned in Sec.~\ref{sec:PEMcoherence}, we can improve the estimate
of the PEM contribution to the coherence 
by allowing for correlations between different PEM channels $\tilde z_I$ 
and $\tilde z_J$. This requires inverting the full matrix of PEM
coherences $\gamma_{IJ}(f)$ or solving a large number of simultaneous
equations involving $\gamma_{IJ}(f)$, rather than simply taking the maximum 
of the product of the coherences as was done here. A computationally-cheaper 
alternative might be to invert a sub-matrix formed from the largest PEM
contributors---i.e., those PEM channels that contribute the most to the 
coherence.

(ii) We can use {\em bicoherence} techniques to account for (non-linear) 
up-conversion processes missed by standard coherence calculations. This may 
allow us to identify cases where low-frequency disturbances excite 
higher-frequency modes in the detector.

(iii) The estimators $\hat\Omega_{\alpha}(f)$ used in this analysis are 
optimal in the {\em absence} of correlated noise. In the presence of 
correlated noise, these estimators are biased, with expected values given by 
the sum, $\Omega_\alpha+\eta_\alpha(f)$, where the second term involves 
the cross-spectrum, $N_{12}(f)$, of the noise contribution to the detector 
output. An alternative approach is to start with a likelihood function for the 
detector output $\tilde s_1$, $\tilde s_2$, where we allow (at the outset) 
for the presence of cross-correlated noise. (This would show up in the 
covariance matrix for a multivariate Gaussian distribution.) We can 
parametrize $N_{12}(f)$ in terms of its amplitude, spectral index, etc.,
and then construct posterior distributions for these parameters along with 
the amplitude and spectral index of the stochastic GW signal. In this 
(Bayesian) approach, the cross-correlated noise is treated on the same 
footing as the stochastic GW and is estimated (via its posterior distribution)
as part of the analysis \cite{robinson-vecchio}. However, as described 
in \cite{bayesnoise}, this works only for those cases where the spectral 
shapes of the noise and signal are different from one another.

(iv) We can also reduce correlated noise by first removing as much noise as 
possible from the output of the {\em individual} detectors. Wiener filtering 
techniques can be applied to remove acoustic, magnetic, and gravity-gradient 
noise from the time-series output of the LIGO 
detectors~\cite{1999gr.qc9083A,actNoiseCancel,driggers-et-al}.
Furthermore, feed-forward control can be used to to cancel seismically-induced 
motion before it affects the LIGO test masses~\cite{actNoiseCancel}.

These and/or other techniques might be needed for future cross-correlation 
searches using advanced detectors, where improved (single-detector) 
sensitivity will mean that correlated noise may be an issue even for 
physically-separated detectors, such as the LIGO Hanford-LIGO Livingston 
detector pair \cite{ChristensenT,wsubtract,schumann}.

\subsection*{Acknowledgment}
The authors gratefully acknowledge the support of the United States
National Science Foundation for the construction and operation of the
LIGO Laboratory and the Science and Technology Facilities Council of the
United Kingdom, the Max-Planck-Society, and the State of
Niedersachsen/Germany for support of the construction and operation of
the GEO600 detector. The authors also gratefully acknowledge the support
of the research by these agencies and by the Australian Research Council,
the International Science Linkages program of the Commonwealth of Australia,
the Council of Scientific and Industrial Research of India, the Istituto
Nazionale di Fisica Nucleare of Italy, the Spanish Ministerio de
Educaci\'on y Ciencia, the Conselleria d'Economia, Hisenda i Innovaci\'o of
the Govern de les Illes Balears, the Royal Society, the Scottish Funding
Council, the Scottish Universities Physics Alliance, The National Aeronautics
and Space Administration, the Carnegie Trust, the Leverhulme Trust, the David
and Lucile Packard Foundation, the Research Corporation, and the Alfred
P. Sloan Foundation.

\bibliographystyle{apsrev}
\bibliography{S5H1H2_paper}

\end{document}